\documentclass[journal=jacsat,manuscript=article]{achemso}

\usepackage[version=3]{mhchem} 
\usepackage{amsmath,amsthm,amssymb,amsfonts}
\usepackage{float}
\usepackage{siunitx}
\usepackage{graphicx}
\usepackage{booktabs}
\usepackage[english]{babel}
\usepackage{textgreek}

\usepackage{hyperref}

\DeclareSIUnit\angstrom{\text {Å}}

\newcommand{\beginsupplement}{%
        \setcounter{table}{0}
        \renewcommand{\thetable}{S\arabic{table}}%
        \setcounter{figure}{0}
        \renewcommand{\thefigure}{S\arabic{figure}}%
        \setcounter{equation}{0}
        \renewcommand{\theequation}{S\arabic{equation}}%
        \setcounter{section}{0}
        \renewcommand{\thesection}{S\arabic{section}}%
        \setcounter{subsection}{0}
        \renewcommand{\thesubsection}{S\arabic{subsection}}%
        \newcounter{SItab}
        \renewcommand{\theSItab}{S\arabic{SItab}}%
        \newcounter{SIfig}
        \renewcommand{\theSIfig}{S\arabic{SIfig}}%
}

\author{Debarshi Banerjee}
\affiliation[ICTP]{International Centre for Theoretical Physics (ICTP), Strada Costiera 11, 34151 Trieste, Italy}
\alsoaffiliation[SISSA]{Scuola Internazionale Superiore di Studi Avanzati (SISSA), via Bonomea 265, 34136 Trieste, Italy}
\author{Sonika Chibh}
\affiliation[TAU1]{The Shmunis School of Biomedicine and Cancer Research, The George S. Wise Faculty of Life Sciences, Tel Aviv University, 6997801 Tel Aviv, Israel}
\author{Om Shanker Tiwari}
\affiliation[TAU1]{The Shmunis School of Biomedicine and Cancer Research, The George S. Wise Faculty of Life Sciences, Tel Aviv University, 6997801 Tel Aviv, Israel}
\author{Gonzalo Díaz Mirón}
\affiliation[ICTP]{International Centre for Theoretical Physics (ICTP), Strada Costiera 11, 34151 Trieste, Italy}
\author{Marta Monti}
\affiliation[ICTP]{International Centre for Theoretical Physics (ICTP), Strada Costiera 11, 34151 Trieste, Italy}
\author{Hadar R. Yakir}
\affiliation[HUJ]{Institute of Chemistry, The Hebrew University of Jerusalem, Edmond J. Safra Campus, 9190401 Jerusalem, Israel}
\author{Shweta Pawar}
\affiliation[BIU]{Faculty of Engineering, The Institute of Nanotechnology and Advanced Materials, Bar-Ilan University, 5290002 Ramat-Gan, Israel}
\author{Dror Fixler}
\affiliation[BIU]{Faculty of Engineering, The Institute of Nanotechnology and Advanced Materials, Bar-Ilan University, 5290002 Ramat-Gan, Israel}
\author{Linda J. W. Shimon}
\affiliation[WEIZ]{Department of Chemical Research Support, Weizmann Institute of Science, 7610001 Rehovot, Israel}
\author{Ehud Gazit}
\affiliation[TAU1]{The Shmunis School of Biomedicine and Cancer Research, The George S. Wise Faculty of Life Sciences, Tel Aviv University, 6997801 Tel Aviv, Israel}
\alsoaffiliation[TAU2]{Department of Materials Science and Engineering, The Iby and Aladar Fleischman Faculty of Engineering, Tel Aviv University, 6997801 Tel Aviv, Israel}
\alsoaffiliation[TAU3]{Sagol School of Neuroscience, Tel Aviv University, 6997801 Tel Aviv, Israel}
\author{Ali Hassanali}
\affiliation[ICTP]{International Centre for Theoretical Physics (ICTP), Strada Costiera 11, 34151 Trieste, Italy}
\email{ahassana@ictp.it}

\title{Turning on the Light: Polymorphism-Induced Photoluminescence in Cysteine Crystals}

\abbreviations{NAF,TSH,NAMD}

\SectionNumbersOn
\begin{document}


\begin{abstract}
Photoluminescence of non-aromatic supramolecular chemical assemblies has attracted considerable attention in recent years due to its potential for use in molecular sensing and imaging technologies.  The underlying structural origins, the mechanisms of light emission in these systems, and the generality of this phenomenon remain elusive. Here, we demonstrate that crystals of L-Cysteine (Cys) formed in heavy water (\ce{D2O}) exhibit distinct packing and hydrogen-bond networks, resulting in significantly enhanced photoluminescence compared to those prepared in \ce{H2O}. Using advanced excited-state simulations, we elucidate the nature of electronic transitions that activate vibrational modes of Cys in \ce{H2O}, particularly those involving thiol (\ce{S-H}) and amine (\ce{C-N}) groups, which lead to non-radiative decay. For the crystal formed in \ce{D2O}, these modes appear to be more constrained, and we also observe intersystem crossing from the singlet to the triplet state, indicating a potentially more complex light emission mechanism. Our findings provide new insights into this intriguing phenomenon and introduce innovative design principles for generating emergent fluorophores.


\end{abstract}

\section{Introduction} \label{sec:intro}
The formation of ordered supramolecular assemblies by various metabolites including amino acids, has been extensively studied in recent years. These simple and biocompatible building blocks can assemble into ordered architectures with unique chemical and physical properties \cite{gupta2021amino,bera2020self,kordasht2021poly,geng2020covalent,zhao2021fabrication}. One of the most intriguing properties within this context, is aggregation-dependent luminescence which is also commonly referred to in the literature as cluster-triggered emission\cite{agg_hong2009aggregation,agg_yang2020organic,agg_liu2020mdm2,agg_zhao2020aggregation,zheng2020accessing,agg_xie2024cluster}. Single amino acids or polypeptide chains in isolation do not display these anomalous optical properties -- however, crystals or aggregates of these monomeric units appear to present exotic absorption and emission spectra\cite{shukla2004novel,arnon2021off,stephens2021short,kumar2022role,2025_chemrxiv_vignesh}. Curiously, this phenomenon appears to contradict the conventional chemical paradigm that fluorescence in chemical and biological systems arises from aromatic or, more generally, conjugated groups \cite{lakowicz2006principles}.

This unexpected non-aromatic fluorescence (NAF) has now been observed in a variety of biological and synthetic supramolecular assemblies constituting amino acids\cite{arnon2021off,kumar2022role,stephens2021short,2025_chemrxiv_vignesh}, peptides\cite{shukla2004novel}, peptide-based dendrimers\cite{lee2004strong,wang2004fluorescence}, oligosaccharides\cite{yu2019oligosaccharides}, amyloid fibrils\cite{del2007charge,pinotsi2016proton,johansson2017label,chung2022label,pansieri2019ultraviolet,pinotsi2013label}, and other proteins\cite{bhattacharya2017direct,prasad2017near,kumar2020weak}, all of which lack aromatic building blocks.  In the last decade, there have been several proposals on the electronic and spectroscopic origins of this phenomenon. This includes the delocalization of electrons across the peptide backbone structure\cite{shukla2004novel}, the existence of short hydrogen-bonds that permit proton transfer\cite{pinotsi2016proton,stephens2021short}, charge transfer excitations between polar amino acids\cite{prasad2017near}, and constrained carbonyl groups leading to fluorescence\cite{niyangoda_2017_plosone,Grisanti_JACS_2020,morzan_2022_jpcb,CO_lock_2023}. The manifestation of these various effects in different types of aggregates or assemblies with different chemical properties, remains an open area of study from both experimental and theoretical perspectives.

Organic crystals offer excellent model systems for exploring the microscopic origins of NAF since one can determine the key interactions governing this phenomenon in a more controlled setting. It is widely recognized that molecular packing plays a crucial role in organic solid-state emission \cite{an2015stabilizing}, where even minor changes in molecular arrangement can lead to substantial variations in photophysical properties \cite{varughese2014non,safin2016polymorphism,yang2018influence,li2019impact}. For example, different crystal polymorphs of the same compound can display variations in emission wavelengths or engage in fundamentally different photophysical processes, including thermally activated delayed fluorescence (TADF), and persistent room temperature phosphorescence (p-RTP) \cite{zhang2010polymorphism,yoon2011polymorphic,gu2012polymorphism,botta2016polymorphism,echeverri2020untangling,deng2025revealing}.  Interestingly, very recent work has shown that changing the chemistry and physical interactions associated with amino acids in crystals through the introduction of sugars, can significantly enhance luminescence\cite{zhang2025_gluser}.

Our search for novel compounds exhibiting NAF, has led us to investigate how crystal polymorphism can affect optical properties in L-Cysteine (Cys). The thiol chemistry associated with Cys is essential for stabilizing tertiary and quaternary protein structures\cite{creighton1988disulphide,welker2002intramolecular,chen2012intramolecular}. Furthermore, the thiol group also makes Cys extremely sensitive to redox chemistry -- a factor that has been linked to the medical pathology of various diseases associated with protein aggregation\cite{puka1995neurotoxicity,janaky2000mechanisms,guttmann2012redox,schafheimer2014tyrosine,valle2017cysteine}. Within the context of fluorescence spectroscopy, Cys has also been suggested to function as a quencher of fluorescence of aromatic and heterocyclic compounds \cite{steiner1969interaction,harris1990photophysics,chen1998toward,qiu2008ultrafast}. 

In this work, we present a joint experimental and theoretical study expanding our understanding of non-aromatic fluorescence focusing on Cysteine crystals grown in \ce{H2O} (Cys-H) and \ce{D2O} (Cys-D). 
Over the last decades, there has been a growing appreciation that biophysical processes\cite{d2o_lehrer1970,d2o_mischa-bonn2023} such as protein folding and aggregation can be subtly modulated using either light (\ce{H2O}) or heavy water (\ce{D2O}). Surprisingly, these two crystals have starkly different crystal structures as well as optical properties. Interestingly, Cys-D exhibits enhanced fluorescence compared to Cys-H. This enhancement originates from subtle differences in the molecular packing and hydrogen-bonding networks of Cys crystals formed in light versus heavy water. By deploying excited-state non-adiabatic molecular dynamics simulations, we unravel the nature of the electronic states that trigger non-radiative decay pathways involving the amine (\ce{C-N}) and thiol (\ce{S-H})  groups differently for Cys structures crystallized in light and heavy water. We also show the importance of the coupling of both singlet and triplet state emission for Cys-D indicating a much more complex origin in the underlying photoluminescence than is evident at first sight. These insights could guide the rational design of bioorganic fluorophores with tunable emission, opening new avenues for imaging, sensing, and photonics in biological systems.

\section{Results} \label{sec:results}

\subsection{Fluorescence of Cysteine in \texorpdfstring{\ce{H2O} vs \ce{D2O}}{H2O vs D2O}} \label{subsec:expt_fluo_results}
We prepared crystals of L-Cys in \ce{D2O} and \ce{H2O} using the heating and cooling method\cite{tiwari2023entropically}. Specifically, L-Cys was dissolved in \ce{H2O} and \ce{D2O} at \SI{90}{\celsius} until the solution became clear and transparent. Subsequently, the amino acid solutions were cooled to room temperature to allow crystals to form.

\begin{figure}[ht!]
\centering
\includegraphics[width=\linewidth]{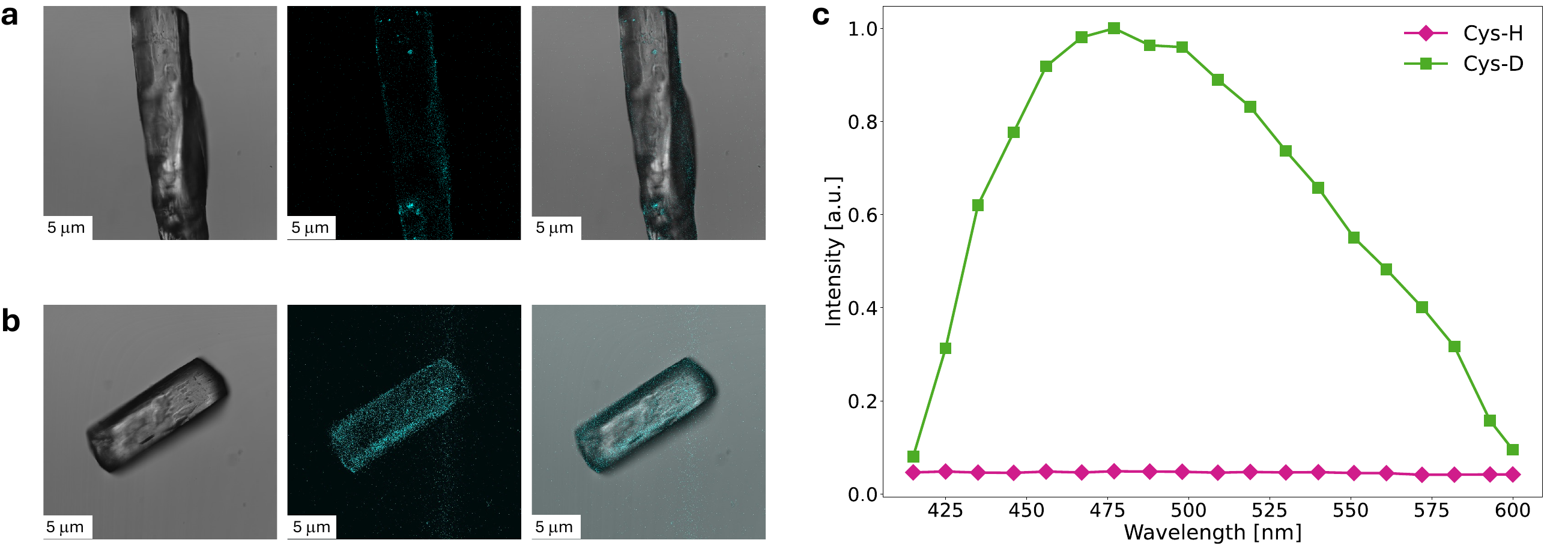}
\caption{In panels \textbf{a} and \textbf{b}, the confocal microscopic images of single crystals of Cys-H and Cys-D that are excited at a wavelength of \SI{405}{\nano\meter} are shown, respectively. Each of these two panels consist of brightfield, fluorescence, and merged image of the crystals (from left to right). In panel \textbf{c}, the emission spectra of the two systems captured with a confocal microscope is shown, with Cys-H in pink, and Cys-D in green.}
\label{fig:expt_spectra}
\end{figure}

Following crystallization, confocal microscopy was used to investigate the optical properties of the crystals. The crystals were illuminated by laser light, and their fluorescence was recorded. An excitation wavelength of \SI{405}{\nano\meter} was used to capture images of fluorescence and brightfield. It was evident from the images taken of the crystals formed in \ce{H2O} that the fluorescence was very low, as seen in Fig.~\ref{fig:expt_spectra}a. Crystals formed in \ce{D2O} however, showed significantly higher fluorescence than crystals formed in \ce{H2O} (Fig.~\ref{fig:expt_spectra}b). Confocal microscopy was also performed on multiple crystals together and the same trend was observed (see SI Figure \ref{fig:SI_expt_flim_agg}a-b).

To obtain a more quantitative comparison of the fluorescence intensity arising from the crystals formed in \ce{H2O} versus \ce{D2O}, the emission spectra were extracted from the confocal microscope and the fluorescence signals were quantified. The crystals prepared in \ce{D2O} produced more enhanced fluorescence signals that were approximately an order of magnitude higher than those prepared in \ce{H2O} (Fig.~\ref{fig:expt_spectra}c). 

Although the preceding results indicate that the optical properties of the crystals in \ce{H2O} and \ce{D2O} are different, the setup of the confocal microscope has a limited sensitivity for resolving the fluorescence intensity. We thus turned to using a time-correlated single-photon counting (TCSPC) system to conduct fluorescence lifetime imaging microscopy (FLIM). Cys crystals were uniformly distributed on a microscope slide prior to imaging. The FLIM system, equipped with a TCSPC card and two detectors, was used to capture images and record the fluorescence lifetimes of the crystals in both \ce{H2O} and \ce{D2O} under identical conditions. Using FLIM, we also find that the crystals in \ce{D2O} display enhanced fluorescence compared to those in \ce{H2O} (see SI Figure \ref{fig:SI_expt_flim_agg}c). In addition, the absolute quantum yield was determined for both Cys-H and Cys-D at room temperature, the former being $0 (\pm 0.05)\%$ (that is, there was no measurable quantum yield within the error bars of the instrument), and the latter at $8 (\pm 0.2)\%$ (see SI Figure \ref{fig:SI_quantum_yield} for details).



\subsection{Crystal Structure of Cysteine in \texorpdfstring{\ce{H2O} vs \ce{D2O}}{H2O vs D2O}} \label{subsec:expt_crys_results}

Cysteine has two well-known crystal polymorphs when crystallized in \ce{H2O} at both ambient and low temperatures. One of them is a monoclinic crystal\cite{cys-h_harding1968crystal,cys-h_crystal_gorbitz1996}, while the other is orthorhombic\cite{cys-h_kerr1973structure,cys-h_kerr1975neutron,cys-h_moggach2005cysteine}. The striking differences between the optical properties of Cys-H and Cys-D indicate that there must be some important structural differences between the two crystals. Therefore, to understand the structural features contributing to the differences in the optical properties between the two crystals in \ce{H2O} and \ce{D2O}, we first investigated their crystal packing. We resolved the latter structure by performing single-crystal X-ray diffraction (SC-XRD) (see Methods for more details), while the former was taken from Ref.\citenum{cys-h_crystal_gorbitz1996}. Crystal data collection and refinement parameters are shown in Supplementary Table \ref{tab:SI_crystal_data}. To the best of our knowledge, this is the first time the structure of Cys has been crystallized and characterized using \ce{D2O} as a solvent. Neither Cys-H nor Cys-D contains any crystallographic light/heavy water molecules. One interesting feature to note is that the Cys-D structure bears a striking similarity to the orthorhombic structures that have been observed in Refs.\citenum{cys-h_kerr1973structure,cys-h_kerr1975neutron,cys-h_moggach2005cysteine,cys_h-moggach2006high}, indicating an underlying crystal polymorphism.


\begin{figure}[ht!]
\centering
\includegraphics[width=\linewidth]{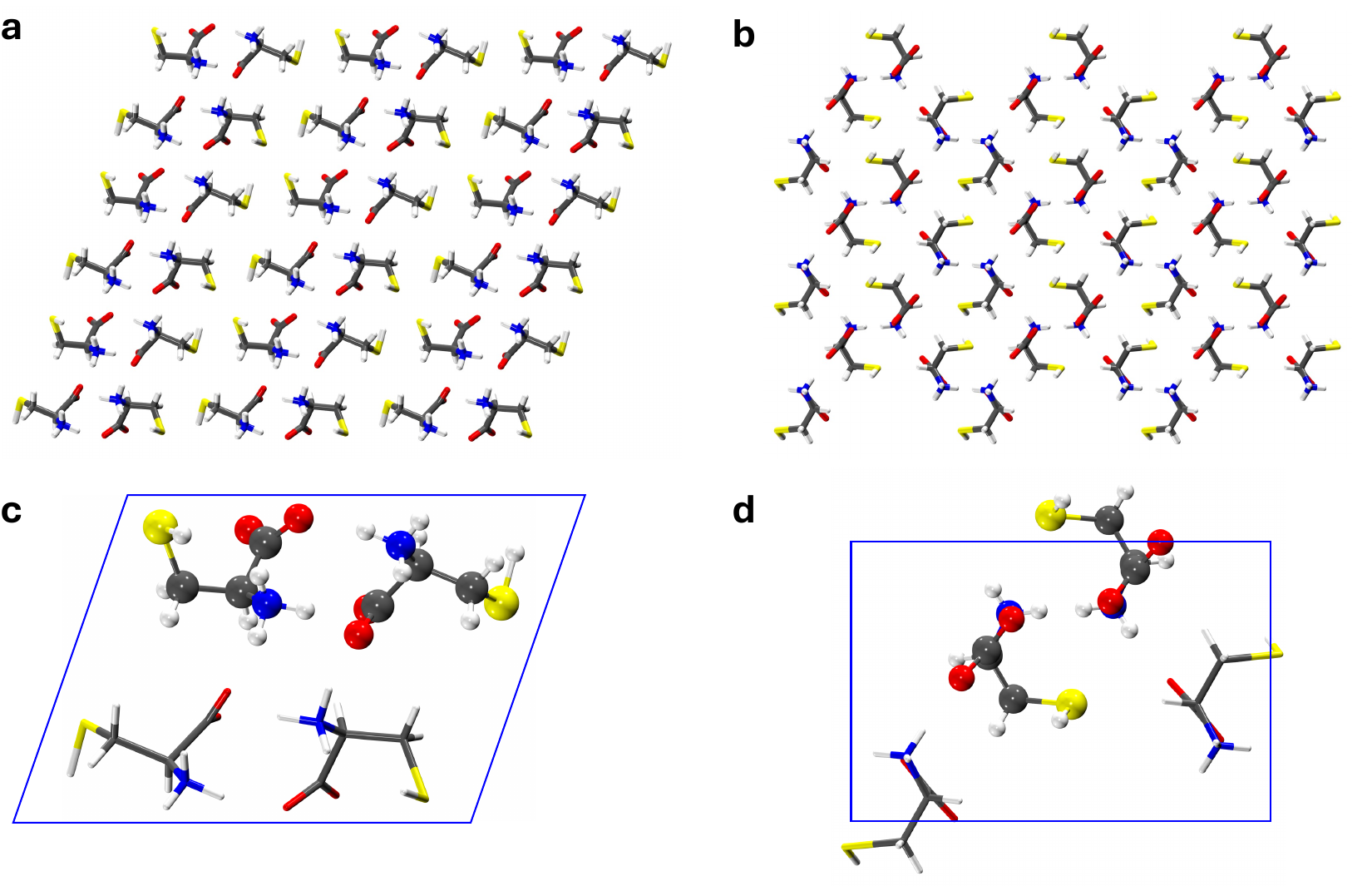}
\caption{Panels \textbf{a} and \textbf{b} show an overview of the crystal packing in Cys-H\cite{cys-h_crystal_gorbitz1996} and Cys-D respectively. Panel \textbf{c} shows the monoclinic unit cell of Cys-H, while \textbf{d} shows the orthorhombic unit cell of Cys-D, each containing 4 monomers, where the QM monomers in a QM/MM framework are shown in ``CPK" representation.}
\label{fig:expt_crystal}
\end{figure}


The two different structures thus formed by crystallizing cysteine in \ce{H2O} (Cys-H) and \ce{D2O} (Cys-D) are shown in Fig.~\ref{fig:expt_crystal}a (monoclinic) and Fig.~\ref{fig:expt_crystal}b (orthorhombic). Both crystals consist of 4 monomers in the unit cell with subtle differences in the packing arrangement, specific interactions involving the hydrogen bonds (HBs), and the conformations of the cysteine molecules (Fig.~\ref{fig:expt_crystal}c-d). Specifically, the N-C termini HBs typically range between 2.70--2.87 \si{\angstrom} in Cys-H while in Cys-D, the N-C termini HBs range between 2.76--2.78 \si{\angstrom}. Similarly, the thiol groups in Cys-H tend to interact with each other through weak HB-like interactions around 3.6 \si{\angstrom}, while in the case of Cys-D these geometrical parameters are longer by $\sim 7\%$ (3.84 \si{\angstrom}). Differences along other dihedral angles involving the \ce{O-C-C_\alpha-C_\beta} are also observed (see SI Figure \ref{fig:SI_OCCaCb_GS}) which ultimately play an important role in affecting the fate of cysteine on the excited state.

While the XRD experiments provide insight into the different crystal structures of Cys-H and Cys-D, they do not reveal the positions of hydrogen atoms, as these cannot be resolved. However, given the subtle differences between Cys-H and Cys-D, it is important to establish whether these crystallographic differences arise from a hydrogen isotope effect or an environmental solvent effect. Therefore, electrospray ionization mass spectrometry (ESI-MS) in positive ion mode was performed on both crystals and the results are shown in SI Figure \ref{fig:SI_mass_spec}. The mass spectrum of Cys-H (SI Figure \ref{fig:SI_mass_spec}a) shows an intense m/z peak at 122.1, which is consistent with the fragment of the protonated molecule ([cySH+H]$^+$)\cite{ESI-MS-ref2012}. The same m/z peak is observed at 122.4 for Cys-D (SI Figure \ref{fig:SI_mass_spec}b), thus confirming the presence of hydrogen atoms in both crystals. Other characteristic peaks of cysteine, i.e. at m/z 152 and 241, can be also observed in the spectra, as well as the ion peak corresponding to sodiated cystine ([cySScy+Na]$^+$) at m/z 263. The detected ion peaks provide further validation that the differences in the physical properties in the two structures do not originate from isotope differences - all non-carbon, non-nitrogen, and non-sulfur atoms are hydrogen in both crystals.



\subsection{Excited-State Simulations: Cys-H vs Cys-D}  \label{subsec:theo_results}
The preceding experimental results show a striking effect of the solvent (\ce{H2O} vs \ce{D2O}) on the resulting cysteine crystal structures. To understand the differences in the optical properties of Cys-H and Cys-D, we turned to conducting mixed quantum/classical molecular (QM/MM) electronic structure calculations that probe the absorption, emission, and non-radiative decay mechanisms. Time-Dependent Density Functional Theory (TDDFT) with the CAM-B3LYP functional was used for all our electronic structure calculations (see Methods for more details). We carved out two monomers from each unit cell forming a dimer which was treated at QM level (see Fig.~\ref{fig:expt_crystal}c-d). Note that the criterion for choosing these QM dimers was based on making sure that the highest number of HBs were formed between the N and C termini. This was then embedded in a periodic environment consisting of MM cysteine molecules. Given the high computational cost of these simulations, we strike a careful balance between efficiency and accuracy, guided by previous approaches introduced by some of us\cite{CO_lock_2023,gonza_jctc_2024_dftb_naf}. In the Methods section, we present validation tests using more accurate electronic structure methods and a larger QM region to justify our approach.

\begin{figure}[ht!]
\centering
\includegraphics[width=\linewidth]{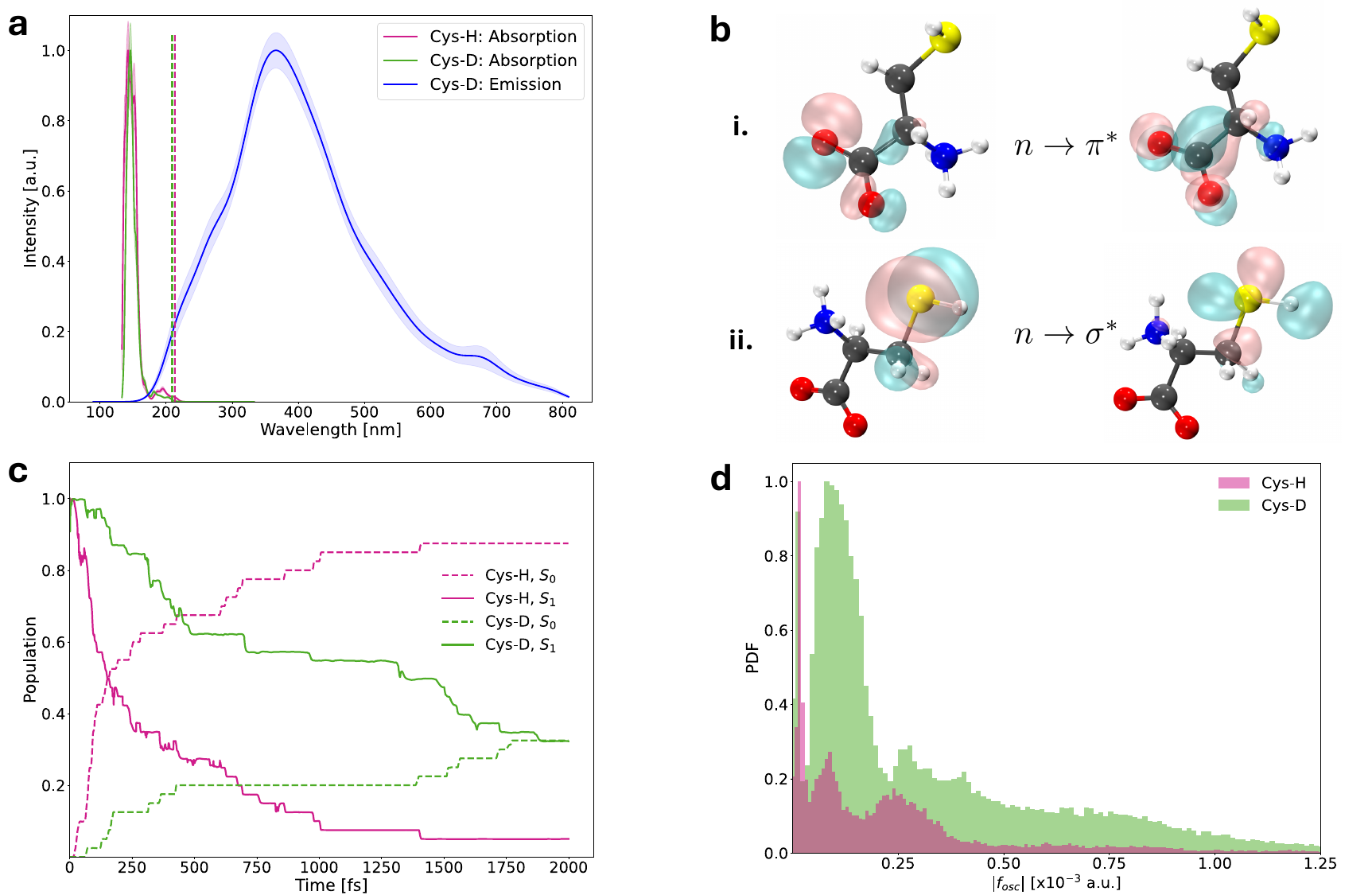}
\caption{In panel \textbf{a}, the theoretically computed absorption spectra for Cys-H (pink) and Cys-D (green) and fluorescence emission spectrum (blue) for Cys-D is shown. The shaded blue region denotes the standard error in theoretical estimation since it is an average over 40 independent trajectories. The dashed vertical lines (green - \SI{209}{\nano\meter}, pink - \SI{215}{\nano\meter}) represents the wavelength corresponding to the $\text{S}_0 \rightarrow \text{S}_1$ transition in our electronic structure calculations for Cys-D and Cys-H, respectively. Panel \textbf{b} shows the 2 types of $\text{S}_0 \rightarrow \text{S}_1$ transitions that occur commonly. In (\textbf{b.i}), we see an $\text{n} \rightarrow \pi^*$ transition that occurs on the carbonyl (CO) group, whereas in (\textbf{b.ii}) we see an $\text{n} \rightarrow \sigma^*$ transition that is present in the SH moiety. Panel \textbf{c} shows the evolution of the average population of the singlet electronic states, where $\text{S}_1$ and $\text{S}_0$ are denoted by solid and dashed lines respectively. Results corresponding to Cys-D and Cys-H are shown in green and pink, respectively. In panel \textbf{d}, the oscillator strengths obtained from the TSH simulations, corresponding to $\text{S}_1 \rightarrow \text{S}_0$ emission are shown. Results corresponding to Cys-D and Cys-H are shown in green and pink, respectively.}
\label{fig:namd_spectra_ensemble}
\end{figure}

We first determined the absorption spectra for both systems using TDDFT, based on a sample of 40 independent configurations from the ground state equilibrium simulations (see Methods section). Figure~\ref{fig:namd_spectra_ensemble}a shows the absorption spectra for Cys-H and Cys-D. In both cases, the lowest excitation energies are found to be around \SI{210}{\nano\meter} (\SI{5.9}{\electronvolt}) (see SI Figure \ref{fig:SI_spectra_theo}a for a zoomed-in version). To gain further insight into the nature of the low-lying electronic transitions, we examined the molecular orbitals involved in the $\text{S}_0\rightarrow \text{S}_1$ excitations. Figure~\ref{fig:namd_spectra_ensemble}b illustrates the molecular orbitals associated with two typical excitations observed in both systems: i) $\text{n} \rightarrow \pi^*$ transition involving the C-terminus carboxylate groups and ii) $\text{n} \rightarrow \sigma^*$ transition localized on the \ce{S-H} (thiol) group.

Although both of these transitions can characterize the $\text{S}_0\rightarrow \text{S}_1$ excitation, there are differences in their occurrence in both systems. In Cys-H, the initial excitation is more or less equally probable to be of either type, while for Cys-D, it is much more likely to be of type i) than type ii) (see Table \ref{tab:excitations} for details). The differences in crystal packing between Cys-H and Cys-D seem to significantly influence the nature of electronic transitions in the Franck-Condon region. Consequently, the variations observed in fluorescence are expected to become more pronounced as the system evolves in its electronically excited states.


Analysis of the 40 simulated trajectories for each system revealed distinct non-radiative decay pathways for Cys-H and Cys-D. In Cys-H, 90\% of trajectories decayed non-radiatively to the ground electronic state, while in Cys-D, this pathway was observed in only 33\% of trajectories. This difference corroborates the experimentally observed enhancement of emission intensity upon the change of packing from the crystal structure of Cys-H to that of Cys-D. Furthermore, Cys-D displays sizable intersystem crossing (ISC), with an additional 33\% of trajectories populating the triplet states. In contrast, ISC was less prevalent in Cys-H, accounting for only 7.5\% of the simulated trajectories. At the end of our \SI{2}{\pico\second} NAMD simulations, only 5\% of trajectories remained in $\text{S}_1$ for Cys-H, while 33\% did so for Cys-D. For the latter, we can use the trajectories that remain on $\text{S}_1$ to estimate the fluorescence emission spectrum which is shown by the blue curve in Figure~\ref{fig:namd_spectra_ensemble}a. This emission spectrum is blue-shifted compared to the experimental spectra by about \SI{80}{\nano\meter} (\SI{0.5}{\electronvolt}) which is within the theoretical error bars of the electronic structure methods we are employing. Later in the paper, we present benchmarks to determine theoretical error bars for our choice of electronic structure method and discuss their implications for comparisons with experimental results.

We begin first by examining the behavior of the excited state dynamics involving the singlet states. Figure~\ref{fig:namd_spectra_ensemble}c shows the average electronic populations of the $\text{S}_0$ and $\text{S}_1$ states for both Cys-H and Cys-D. Here we clearly observe that in Cys-H, the $\text{S}_1$ population reduces significantly within the first \SI{250}{\femto\second}, and thereafter continues decaying further, while for Cys-D, the population decay occurs over a significantly longer time-scale. The corresponding curves for the $\text{S}_0$ state are also shown mirroring the behavior of the $\text{S}_1$. These theoretical predictions are thus fully consistent with the experimental observation of photoluminescence only in Cys-D. 

The lower probability of non-radiative decay in Cys-D, on its own, is not sufficient to lead to non-aromatic fluorescence. Specifically, we also need the excited state on which the system exists to be \emph{bright}. This is characterized in terms of the oscillator strength between $\text{S}_1$ and $\text{S}_0$. In Figure~\ref{fig:namd_spectra_ensemble}d we see the probability distribution function of the oscillator strengths for the $\text{S}_1 \rightarrow \text{S}_0$ transition, and we can see clearly that Cys-D has a higher probability of having a larger oscillator strength than Cys-H, thereby contributing to a \emph{bright} photoluminescence which is, again, in good agreement with the experimental results.

\begin{figure}[ht!]
\centering
\includegraphics[width=\linewidth]{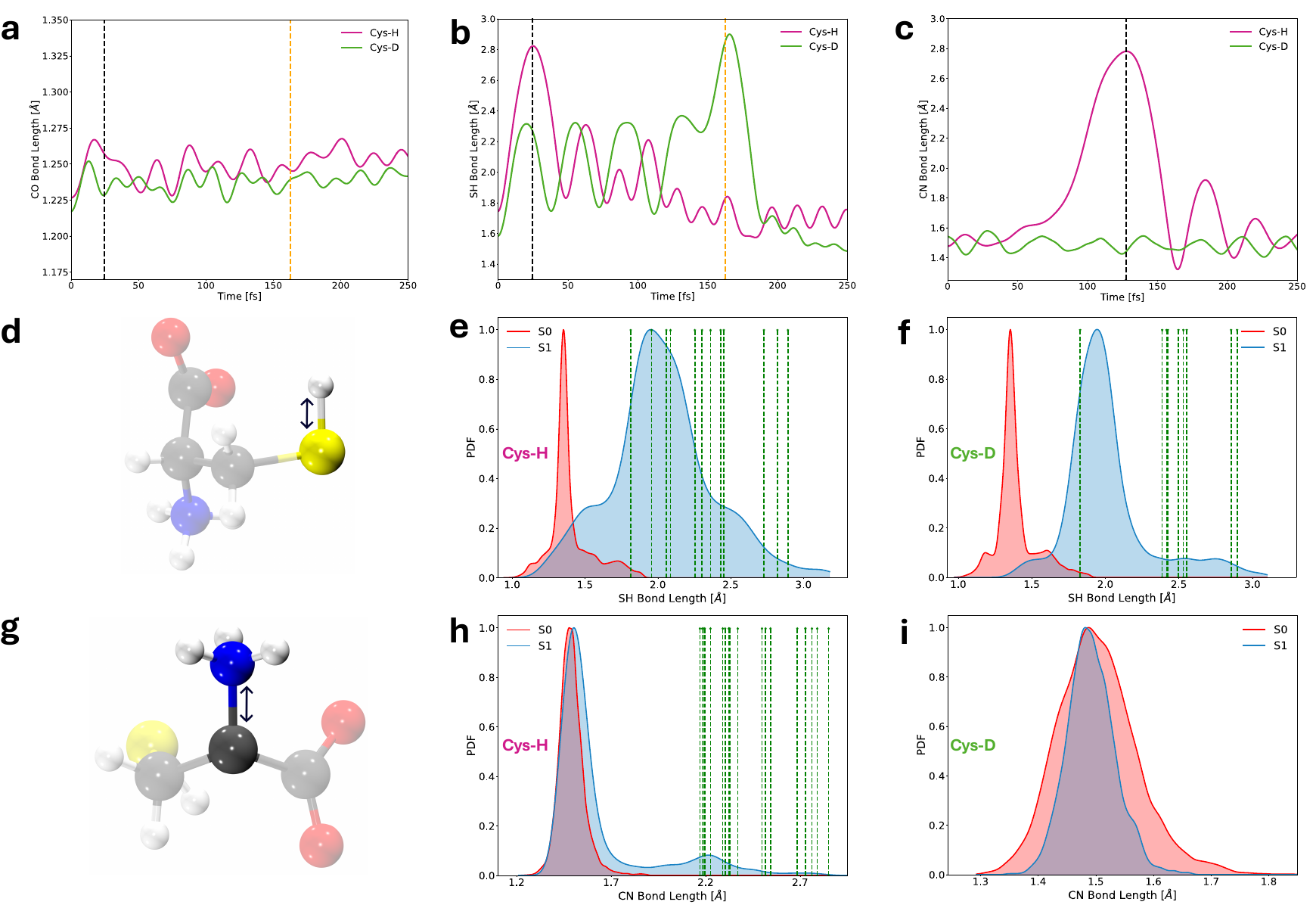}
\caption{Panels \textbf{a},\textbf{b},\textbf{c} show the time-series evolution of CO, SH, CN bond lengths across model trajectories. The bond lengths for Cys-D and Cys-H are shown in green and pink respectively, while the time of encountering the conical intersection is shown with orange and black dashed vertical lines, respectively. The 2 major non-radiative decay modes are the stretching of the \ce{S-H} and \ce{C-N} bonds, depicted in panels \textbf{d} and \textbf{g}. The Probability Distribution Function (PDF) of the \ce{S-H} bond for Cys-H and Cys-D are respectively shown in panels \textbf{e} and \textbf{f}, where the PDF corresponding to $\text{S}_0$ is shown in red and the one corresponding to $\text{S}_1$ is shown in blue. The dashed green lines depict the bond lengths at which the conical intersection is reached, and non-radiative decay occurs from $\text{S}_1 \rightarrow \text{S}_0$. Similarly, in panels \textbf{h} and \textbf{i}, the PDF of the \ce{C-N} bond for Cys-H and Cys-D are shown.}
\label{fig:decay_modes}
\end{figure}

Having established the differences in the energetic and electronic properties on the excited state of Cys-H and Cys-D we next dissect the vibrational modes that lead to non-radiative decay in both systems. In recent works\cite{pinotsi2016proton,Grisanti_JACS_2020,stephens2021short,CO_lock_2023}, research by some of us has shown that specific vibrational modes such as the carbonyl stretch, proton transfer, and deplanarization play a critical role in modulating the excited-state lifetimes of glutamine, peptide chains, and amyloids. In particular, by constraining/locking these modes the barriers to accessing conical intersections (CIs) can be raised. As a result, this pathway is either delayed or completely inhibited, reducing the rate of the non-radiative decay and thereby increasing the likelihood of photon emission. The excitations involving the n-$\pi^*$ and n-$\sigma^*$ shown earlier in Figure~\ref{fig:namd_spectra_ensemble}b, prompted us to examine vibrational coordinates associated with these regions of the molecules. Specifically, in the case of the n-$\pi^*$ transition, a natural expectation would be the lengthening of the \ce{C=O} bond while for the n-$\sigma^*$, there would be an increase in the \ce{S-H} bond length. Table \ref{tab:decay_modes} summarizes the statistics associated with the four main vibrational modes we identified that play a role in the photochemistry, details of which are now discussed.

A visual inspection of some of the excited-state trajectories indicated that most of the activity involves two vibrational modes, the \ce{S-H} and \ce{C-N} bond lengths. Although the initial excitation originates in an n-$\pi^*$ transition localized on the carbonyl group, the \ce{C=O} bond length in all our trajectories displays only moderate fluctuations ranging between 1.1--1.4 \si{\angstrom} and it therefore does not appear to be a key player in the deactivation mechanism. On the other hand, for the excitations involving n-$\sigma^*$ near the sulfide group, the \ce{S-H} bond undergoes large geometrical changes which ultimately lead to non-radiative decays in both Cys-H and Cys-D.

Investigating a sample trajectory for both Cys-H and Cys-D that undergoes non-radiative decay, we show the time-series of the \ce{C=O} bond (Fig.~\ref{fig:decay_modes}a) and observe no significant distortions that coincide with the system encountering the CI and decaying non-radiatively to the ground state. Instead, in the time series of the \ce{S-H} bond (Fig.~\ref{fig:decay_modes}b) we observe a large extension of this bond at the point the system decays to $\text{S}_0$. For Cys-H, 12 out of the 35 decays to the ground state occur through this mechanism, whereas for Cys-D, 9 out of 13 non-radiative decays to the ground state follow this pathway. 

From the preceding analysis, the majority of the non-radiative decays for Cys-H remain unexplained and this led to us looking for alternative decay modes. In Fig.~\ref{fig:decay_modes}c we show a different trajectory where the \ce{C-N} stretches to a large value (\SI{2.8}{\angstrom}) in concert with a decay of the system to the ground state. In fact, this turns out to be the dominant mode of non-radiative decay to the ground state for Cys-H, with 19 out of the 35 decays following this mechanism. Remarkably, this mode of decay appears unique to Cys-H and does not occur in Cys-D. To conclude the discussion of decay mechanisms, the remaining 4 out of 35 decays for Cys-H were through the stretching of the \ce{C-S} bond, often involving a dissociation of the same. For Cys-D, there was 1 out of the 13 decays involving a \ce{C-S} stretch. The remaining 3 decays for Cys-D were through a stretch of the \ce{C-COO-} bond, but these decays, which followed ISC, will be elaborated on in Section \ref{subsec:triplets}.

To better quantify the differences in the ground and excited state distributions along some of these coordinates and how they affect the access to the CI, we show probability distributions of the \ce{S-H} and \ce{C-N} modes in the ground and excited state for Cys-H and Cys-D in Figure~\ref{fig:decay_modes}. Looking at the \ce{S-H} mode (visualized in Fig.~\ref{fig:decay_modes}d), we clearly see that in both Cys-H (Fig.~\ref{fig:decay_modes}e) and Cys-D (Fig.~\ref{fig:decay_modes}f) the probability distribution of the \ce{S-H} bond when the system is in $\text{S}_1$ is peaked around 2.0 \si{\angstrom}, while in $\text{S}_0$ the equilibrium bond lengths are usually between 1.2--1.5 \si{\angstrom}. The bond lengths at which the system encounters the CI are marked through green vertical lines, and they range between 2--3 \si{\angstrom}. The position of the CI appears to be closer to the peak of the distribution of the $\text{S}_1$ state in Cys-H compared to Cys-D, consistent with the idea of a less constrained mode in the former as observed in an earlier study for the carbonyl stretch\cite{CO_lock_2023}.

Looking at the probability distribution of the \ce{C-N} mode (visualized in Fig.~\ref{fig:decay_modes}g) in both Cys-H (Fig.~\ref{fig:decay_modes}h) and Cys-D (Fig.~\ref{fig:decay_modes}i), we clearly see that the bond can stretch to larger values in $\text{S}_1$ only in Cys-H, while in Cys-D it remains confined between 1.4--1.6 \si{\angstrom}. This leads to the system decaying through the conical intersection along the \ce{C-N} mode only in Cys-H, and the enhanced optical activity of Cys-D can be primarily attributed to the absence of this mode of decay. We will elaborate more on the \ce{C-N} mode and its uniqueness in Cys-H in Section \ref{subsec:cn_discussion}. Furthermore, the large overlap in the distribution of the \ce{C-N} mode in Cys-D creates a situation that makes accessing the CIs along this mode more difficult.

\subsection{Singlet to Triplet Intersystem Crossing (ISC) in Cys-D} \label{subsec:triplets}

\begin{figure}[ht!]
    \centering
    \includegraphics[width=\linewidth]{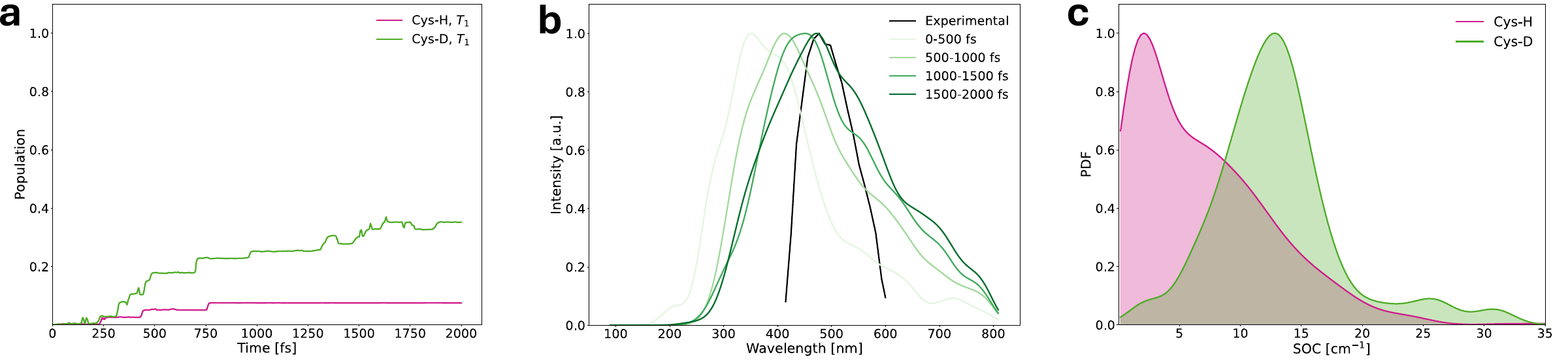}
    \caption{Panel \textbf{a} shows the cumulative triplet state population of Cys-H (pink) and Cys-D (green). In panel \textbf{b} we see the time-dependent emission spectra from the triplet state for Cys-D, in blocks of \SI{500}{\femto\second}, with the shades of green getting darker for successive blocks of time, while the experimental emission spectra is shown in black. Panel \textbf{c} shows a PDF of the SOC computed from 1000 random frames for both Cys-H (pink) and Cys-D (green).}
    \label{fig:triplets}
\end{figure}

As alluded to earlier, in Cys-D, we observe that the photochemistry is also heavily influenced by intersystem crossing. Figure \ref{fig:triplets}a shows the time-evolution of the population of the triplet state for both Cys-H and Cys-D where one clearly sees that the significant increase in the triplet population appears to be a unique feature of Cys-D, where about 35\% trajectories remain in a triplet state at the end of our \SI{2}{\pico\second} dynamics, while 7.5\% of trajectories for Cys-H do the same. In fact, 17 out of 40 trajectories undergo ISC for Cys-D, and 3 of these decay to the ground state from the triplet states. While no clear vibrational mode could be associated to the $\text{S}_1 \rightarrow \text{T}_1$ ISC transitions, the $\text{T}_1 \rightarrow \text{S}_0$ transitions can be associated with a clear stretch of the \ce{C-COO-} bond. This also appears exclusively in Cys-D, and the few trajectories that do undergo ISC for Cys-H do not decay to the ground state via a further ISC.

Our simulations indicate that the ISC in cysteine crystals occurs on a sub-picosecond timescale. Several previous theoretical and experimental studies in a wide-variety of systems have shown \cite{damrauer1997femtosecond,yoon2006direct,ultrafast_isc_jpca_2008,atkins2017trajectory,ultrafast_isc_jacs_2018,ultrafast_isc_molecules_2022} that intersystem crossing can occur on similar timescales. This suggests that the observed luminescence observed on the nanosecond timescale in the experiments may arise from a mix of both singlet and triplet emission. In order to provide a theoretical prediction of the possible emission spectra arising from the lowest lying triplet state, the $\text{T}_1 \rightarrow \text{S}_0$ energy gap, the corresponding transition-dipole moment, and oscillator strength need to be determined. We do this on a subset of configurations, selecting 100 random frames for each \SI{50}{\nano\meter} block in the emission energies between \SI{200}{\nano\meter} to \SI{800}{\nano\meter}. This gives us mean oscillator strengths for each range of energies which we then use to compute the time-dependent emission spectra in blocks of \SI{500}{\femto\second} (Fig.~\ref{fig:triplets}b). Here we observe that the peak of the emission energies appears to converge, perhaps fortuitously, to the peak of the experimental emission spectra. On the other hand, if we look at the time-dependent emission spectra from $\text{S}_1$, it converges to approximately \SI{400}{\nano\meter} which is an \SI{80}{\nano\meter} (\SI{0.5}{\electronvolt}) blue-shift compared to the experimental results (see SI Figure \ref{fig:SI_spectra_theo}b). It is important to note here that the oscillator strengths for triplet emission ($\sim10^{-6}$) are about three orders of magnitude smaller than those corresponding to the singlet emission ($\sim10^{-3}$). Consequently we would expect any emission arising from the triplet states to be much weaker than the singlet emission.

Since the vast majority of the ISC occurs from $\text{S}_1 \rightarrow \text{T}_1$, we look at the likely values of the spin-orbit coupling (SOC) for the two systems in Figure \ref{fig:triplets}c. Having computed the PDF of the SOC over 1000 randomly selected frames while the system is in $\text{S}_1$, we see a clear difference in the two systems, with Cys-D (green) much more likely to have higher values of the SOC than Cys-H (pink), which is consistent with the behavior we observe in our simulations. The significant ISC observed in Cys-D can be further quantified, by visualizing for all trajectories, the cumulative probability of ISC while the system is in $\text{S}_1$. In SI Figure \ref{fig:SI_isc_prob_energy-gaps}a, we clearly see that trajectories of Cys-D have a much higher cumulative probability of ISC than Cys-H. As we have already noted above, Cys-D is more likely to have a higher SOC than Cys-H, and this contributes to the increased ISC probability for the former. Part of the reason might also be the fact that Cys-H undergoes non-radiative decay to $\text{S}_0$ so quickly that the system has no opportunity to experience ISC. Another way of understanding the preferential ISC experienced by Cys-D is by looking at the distribution of energy gaps between $\text{S}_1$ and the closest triplet state\cite{mai2019influence} (which is usually $\text{T}_1$) in SI Figure \ref{fig:SI_isc_prob_energy-gaps}b. In Cys-H, ISC from the singlet excited state is less probable due to the fact that there is a larger gap to the energetically closest triplet state. The distribution has a larger peak near 0 for Cys-D than for Cys-H, which coincides with more frequent occurrences of nearly-degenerate $\text{S}_1$ and $\text{T}_1$ states during Cys-D trajectories. 




\subsection{Inhibiting \texorpdfstring{\ce{C-N}}{C-N} Distortions in Cysteine Luminescence} \label{subsec:cn_discussion}

We have shown that the key modes leading to the non-radiative decay are the stretch of the \ce{S-H} and \ce{C-N} bonds. Extending the \ce{S-H} bond leads to easy access of the conical intersection in both Cys-H and Cys-D. On the other hand, the \ce{C-N} mode is found exclusively in Cys-H. The differences between the two systems originate from a combination of subtle characteristics induced by the conformation of the cysteine molecules in each crystal and their local environment. 

We first examined the evolution of the electronic structure of the $\text{S}_1$ state during a typical decay event. The vast majority of trajectories undergoing a non-radiative decay via the \ce{C-N} stretch, start out with an initial $\text{n}-\pi^*$ transition on the \ce{C=O} at the point of photo excitation (SI Figure \ref{fig:SI_CN_analysis}a.i). Examining the transitions across the trajectory, we notice that the electron density delocalizes along the \ce{C-C_\alpha-C_\beta} chain, and this simultaneously gives rise to a partial $\text{n}-\sigma^*$ on the \ce{C_\alpha-N} bond. These changes make the \ce{C_\alpha-N} bond weaker allowing it to stretch more easily (SI Figure \ref{fig:SI_CN_analysis}a.ii). 

These subtle changes in the underlying electronic structure are ultimately rooted in the differences of the \ce{O-C-C_\alpha-C_\beta} dihedral angles between Cys-H and Cys-D. Specifically, looking at the probability distribution of this dihedral while the system is in $\text{S}_1$ (see SI Figure \ref{fig:SI_CN_analysis}b), we see that a planar geometry (\SI{0}{\degree}, \SI{180}{\degree}, \SI{360}{\degree}) is far more probable in Cys-H and effectively not sampled in Cys-D. This means that the extended \ce{O-C-C_\alpha-C_\beta} planarization is likely to lead to an enhancement of resonance within the molecule in the former.

Another feature that plays an important role is local structural differences between the two crystals. In particular, there appear to be a larger number of oxygen atoms around the N-terminus in Cys-H compared to Cys-D (see the N-O running coordination number function in SI Figure \ref{fig:SI_CN_analysis}c)). This negative electrostatic potential around the N-terminus serves as a trigger to facilitate dissociation of the C-N bond. This feature is also reflected in potential energy surface scans along the C-N bond which show a barrier of over 1 eV for dissociation in the $\text{S}_1$ state in Cys-D compared to Cys-H (SI Figure \ref{fig:SI_CN_analysis}d).

\section{Conclusions} \label{sec:conclusion}

In this joint experimental and theoretical work, we have taken another step to expand our understanding of anomalous fluorescence arising from non-aromatic systems. Cysteine molecules crystallized in light and heavy water display rather striking differences in their crystallographic packing and optical properties. Using excited-state simulations, we rationalize the electronic and vibrational origins of the enhanced fluorescence observed in the crystal that is formed in heavy water. 

The ability to probe in a time-resolved manner nucleation mechanisms of amino acids or proteins in solution remains an open challenge. The intrinsic fluorescence of amino-acids and how they potentially change as a function of clustering or aggregation offers a non-invasive manner to study both thermodynamics and kinetic pathways of crystallization without the introduction of external fluorescent probes. These findings offer exciting perspectives for new design principles to engineer biological systems with enhanced optical properties. Previous research has pointed to the importance of strong hydrogen bonds that facilitate proton transfer\cite{pinotsi2016proton,stephens2021short}, constrained carbonyl bonds\cite{niyangoda_2017_plosone,Grisanti_JACS_2020,morzan_2022_jpcb,CO_lock_2023} and charge-transfer excitations between polar amino acids\cite{prasad2017near}. Here we have deepened our understanding of the problem by elucidating the importance of nominally weaker interactions such as thiol groups (\ce{S-H}) as well as vibrational modes involving the amide backbone (\ce{C-N}).

Recent studies have highlighted that incorporating heavy atoms, such as sulfur, can improve the efficiency of thermally activated delayed fluorescence (TADF) photocatalysts \cite{hojo2023imidazophenothiazine}. Our findings indicate that Cys-D, with its relatively high SOC, could open new possibilities for the design of such materials. However, further experiments, beyond the scope of this work, may be required to validate this direction.

Solvent isotope effects in biophysical processes such as altering the secondary structure and aggregation of proteins have been observed in numerous systems. Here we have shown that the simple effect of moving from light to heavy water alters the crystallization pathways and ultimately also the optical properties of the two crystals. The molecular origins of these effects and how sensitive they are to the chemistry and size of the underlying polypeptide remains an open question.



In summary, this study establishes a fundamental principle of non-aromatic fluorescence which is a clear testament to how relatively subtle structural packing changes can lead to massively different optical properties for crystal polymorphs of amino acids. Based on this molecular picture, we may be able to rationally design new generations of non-aromatic amino-acid based optical devices of unique photonic and electronic properties as well as intrinsic biocompatibility.

\section{Methods} \label{sec:methods}
\subsection{Experimental Methods} \label{subsec:expt_methods}

\subsubsection{L-Cysteine crystallization}
A solution of L-Cysteine (200 mg) in \SI{1}{\milli\liter} of \ce{H2O} and \ce{D2O} was prepared. To obtain a clear, transparent solution, the samples were heated to \SI{90}{\celsius}. After heating, the solutions were gradually cooled to room temperature to facilitate self-assembly and crystallization.

\subsubsection{Confocal Microscopy}
A Leica SP8 Lightning confocal microscope with Leica Application Suite X (LAS X) software was used to capture confocal images of crystals. A \SI{405}{\nano\meter} laser was used to excite the samples. A wavelength range of \SI{415}{\nano\meter} to \SI{600}{\nano\meter} was set for emission. Image analysis was performed using a Leica SP8 confocal microscope to extract the emission graph.

\subsubsection{Quantum Yield}
Fluorescence measurements were carried out with a Horiba Scientific Fluoromax-4 spectrofluorometer. Absolute fluorescence quantum yield measurements were performed using a Quanta-Phi integrating sphere connected to the Fluoromax-4.

\subsubsection{Fluorescence lifetime microscopy (FLIM)}
All fluorescence lifetime (FLT) measurements were performed using a two-channel laser scanning confocal microscope (DCS 120, Becker \& Hickl GmbH, Berlin, Germany). The full width at half maximum (FWHM) of the excitation pulse for this system is of the order of 10--100 ps. For this study, 256$\times$256 pixel sample areas were excited with \SI{50}{\mega\hertz} \SI{473}{\nano\meter} laser pulses and detected using a \SI{495}{\nano\meter} LP and a 620/60 nm HQ filter and 0.5--2.0 mm pinhole. Each sample was measured at least 3 times for \SI{240}{\second} each measurement. Fluorescence emission spectra were performed using a Cary Eclipse Fluorescence Spectrophotometer (Varian, USA), and were obtained using an excitation of \SI{480}{\nano\meter} and a gain of \SI{400}{\volt}.

\subsubsection{Mass Spectrometry}
The samples for mass spectrometry were prepared by dissolving Cys at a concentration of \SI{200}{\milli\gram\per\milli\liter} in \ce{D2O} and \ce{H2O} separately. This solution was heated to \SI{90}{\celsius} and then allowed to cool gradually. Mass spectrometry analysis was carried out using an Acquity UPLC system connected to a TQD XEVO triple quadrupole ESI source mass spectrometer (Waters, Milford, MA, USA). The measurements were performed in positive ionization mode using ESI-MS.

\subsubsection{Single-crystal X-Ray Diffraction}

\paragraph{Preparation of crystal data collection}
Crystal structures of the Cys molecule were obtained from the Single crystal X-ray diffraction (SCXRD) technique. The following method was used to obtain the crystal structure:

Crystals used for data collection were grown using the slow solvent evaporation method. The dry Cys amino acid powder was first dissolved in \ce{D2O} at concentrations of \SI{200}{\milli\gram\per\milli\liter}. Then, the samples were heated at \SI{90}{\celsius} for 3 hrs followed by a vortex to dissolve them completely and allowed to cool down gradually. The formation of a needle-like crystal of Cys took place after a few days by slow evaporation of the \ce{D2O} solvent. For data collection, crystals were coated in Paratone oil (Hampton Research), mounted on a MiTeGen cryo-loop, and flash-frozen in liquid nitrogen. Single crystal diffraction data were collected at \SI{120}{\kelvin} on a Rigaku XtaLAB diffractometer system with Dectris Pilatus 3R 200K-A detector using CuK$\alpha$ radiation at $\lambda = \SI{1.54184}{\angstrom}$.

\paragraph{Crystal data processing and structural refinement}
The diffraction data were collected and processed using the CrysAlisPro 1.171.42.67a suite of programs (RigakuOD 2022). The structures were solved by direct methods using SHELXT-2016/4 and refined by full-matrix least-squares against F2 with SHELXL-2016/4. Atoms were refined independently and anisotropically, with the exception of hydrogen atoms, which were located in the electron density and  refined isotropically. Crystal data collection and refinement parameters are shown in Supplementary Table \ref{tab:SI_crystal_data}, and the complete data can be found in the cif file as Supporting Information. The crystallographic data have been deposited in the Cambridge Crystallographic Data Centre (CCDC) with ID 2358399 for Cys.


\paragraph{Accession Codes}
The CCDC database under accession code 2358399 contains the supplementary crystallographic data of the Cys crystals in \ce{D2O} for this paper.

\subsection{Theoretical Methods} \label{subsec:theo_methods}

\subsubsection{Classical Molecular Dynamics}
The system was first prepared by extracting the crystal structures and building an appropriately large supercell (6$\times$9$\times$5 for Cys-H and 9$\times$6$\times$5 for Cys-D). 
The Generalized Amber Force Field (GAFF) \cite{gaff_wang2004} was used for all MM parameters. First, the system was equilibrated at \SI{300}{\kelvin} in the NVT ensemble for \SI{10}{\nano\second} with a time-step of \SI{2}{\femto\second} and with the Stochastic Velocity Rescaling (SVR) thermostat\cite{bussi2007canonical} with $\tau = \SI{1.0}{\pico\second}$. Next, a \SI{200}{\nano\second} NVT simulation was performed with the same thermostat parameters and a constraint on all bonds involving hydrogen atoms. From this simulation, 40 independent frames were extracted every \SI{5}{\nano\second}. These frames are used as input for the QM/MM simulations that follow. All the classical simulations were running using AMBER 2023 software\cite{amber2005,ambertools2023}.

\subsubsection{Ground State QM/MM}
To get an accurate idea of the optical properties of these crystals, we need to consider the effect of the environment. Since it is computationally prohibitive to model the dynamics of the entire crystal with a fully quantum-mechanical approach, we use a QM/MM\cite{qmmm_warshel_levitt_1976,qmmm_brunk2015mixed} setup. First, we take the 40 independent frames we extracted from the classical MD, and then select the relevant QM dimers (see Fig.~\ref{fig:expt_crystal}c,d). Then we proceed with the equilibration of the temperature of the selected QM dimer region at \SI{300}{\kelvin}. This is done using ORCA 5.0.4\cite{orca_neese2012,orca_neese2018,orca_neese2020,orca_neese2022,orca_neese2023shark} for the QM region, together with AMBER 2023\cite{amber2005,ambertools2023} for the MM. We work at the DFT level of theory with the CAM-B3LYP functional\cite{camb3lyp_yanai2004} and the Karlsruhe def2-SVP basis set\cite{def2_weigend2005,def2_hellweg2015}. To speedup calculations, the RIJ-COSX\cite{rijcosx_neese2009} approximation is used to estimate the HF exchange and Coulomb integrals. The QM region is equilibrated in the NVT ensemble at \SI{300}{\kelvin}, employing the SVR thermostat\cite{bussi2007canonical} with a very tight coupling ($\tau = \SI{0.01}{\pico\second}$), with no bond constraints, and a time-step of \SI{0.5}{\femto\second}. After \SI{1}{\pico\second}, we verify temperature equilibration and then run it for another \SI{5}{\pico\second} with a more typical thermostat coupling ($\tau = \SI{0.5}{\pico\second}$). Now, the final frames (nuclear positions and velocities) of these equilibrated trajectories are taken as the initial conditions for the NAMD QM/MM simulations that follow. These frames are also used to perform absorption spectra calculations which give us an insight into the nature and energies of relevant transitions.

\subsubsection{Non-Adiabatic Molecular Dynamics}
To study the excited state dynamics and photophysical relaxation of these two systems, we use the Trajectory Surface Hopping (TSH) technique, in particular, the SHARC approach\cite{sharc_richter2011,sharc_mai2018} (surface hopping including arbitrary couplings), which is an extension of Tully's Fewest Switches Surface Hopping (FSSH)\cite{tsh_tully1990molecular} method. This class of methods allow the nuclei to be propagated classically on quantum-chemical potential energy surfaces calculated on-the-fly. For a more detailed overview of TSH, the reader is referred to detailed reviews\cite{tsh_barbatti2011nonadiabatic,malhado2014non,persico2014overview,tsh_crespo2018recent}. 

The advantage of the SHARC formalism is the ease of incorporating both time-dependent non-adiabatic coupling (TD-NAC) and spin-orbit coupling (SOC) on the same footing. We include the first 5 singlet states ($\text{S}_0 - \text{S}_4$) and 5 triplet states ($\text{T}_1 - \text{T}_5$) in the calculation. Notably, in SHARC, the individual $M_s$ components of the triplet states are explicitly included, resulting in a total of 20 electronic states considered in the simulations.

All trajectories were started from $\text{S}_1$ after simulating vertical excitation, and then evolved with a time-step of \SI{0.5}{\femto\second} in the NVE ensemble using the velocity verlet integration scheme\cite{verlet1967computer,swope1982computer}.  The functional and basis set used for NAMD are the same as in the ground-state QM/MM dynamics and excitation energies and oscillator strengths were calculated with TDDFT\cite{tddft_casida1995} using the Tamm-Dancoff Approximation (TDA)\cite{tda_hirata1999} to overcome well-know problems with triplet instabilities\cite{triplets_peach2011,triplets_peach2012}. The usage of TDA in TDDFT has also been shown to improve its behavior at/near conical intersections for a variety of systems\cite{triplets_tapavicza2008mixed,triplets_yang2016conical,triplets_matsika2021electronic}. 

The time-dependent non-adiabatic coupling is computed using the numerical differentiation scheme suggested by Hammes-Schiffer and Tully\cite{tsh_hammes1994proton} which use the overlap integrals between the wavefunctions at different time-steps. In addition, we employed the energy-based decoherence correction proposed by Granucci et al.\cite{tsh_granucci2007critical} with a decoherence parameter of 0.1 Hartree. For the integration of the electronic equation of motion, 25 sub-steps were chosen. The trajectories were propagated for \SI{2}{\pico\second}, with the criteria for a hop to the ground state set at an energy difference between the active state and $\text{S}_0$ at \SI{0.2}{\electronvolt} in accordance with other studies using TDDFT\cite{plasser2014surface,wen2023excited,ibele2020excimer}. In our approach, we adjust the kinetic energy of the system by rescaling the velocity after a surface hop to conserve energy, while no action is taken in case a hop is frustrated due to insufficient kinetic energy.

The TSH simulations were done using a framework of SHARC 3.0.1\cite{SHARC3.0} (TSH implementation), ORCA 5.0.4\cite{orca_neese2012,orca_neese2018,orca_neese2020,orca_neese2022,orca_neese2023shark} (QM calculations), and TINKER 6.3.3\cite{tinker_ponder2004} (MM calculations).

\subsubsection{Validation of Computational Methods}

One important aspect to highlight is that our calculated excitation energy, corresponding to the transition to the first excited state, exhibits a blue shift of 195 nm (\SI{2.84}{\electronvolt}) compared to the experimental value. A potential source of this discrepancy is the choice of functional and basis set\cite{laurent2013td}. However, we validated our approach by testing different combinations of basis sets and functionals for the TDDFT calculations, as well as the algebraic diagrammatic construction to second-order (ADC(2)) method\cite{schirmer1982beyond_adc,trofimov1995efficient_adc2}. As we show in the SI, our def2-SVP/CAM-B3LYP approach is consistent with a variety of other methods including def2-SVP/wB97X-D3 (SI Figure \ref{fig:SI_abs_validation}a), def2-QZVPP/CAM-B3LYP (SI Figure \ref{fig:SI_abs_validation}b), def2-SVP/ADC(2) (SI Figure \ref{fig:SI_abs_validation}c). In all of these cases, the lowest excitation energies corresponding to $\text{S}_0 \rightarrow \text{S}_1$ transitions are relatively similar for both systems. The nature of the transitions also largely remain the same, and are dominated by the $\text{n} \rightarrow \pi^*$ on the \ce{C-O} and $\text{n} \rightarrow \sigma^*$ on the \ce{S-H}. In prior work done in our group\cite{CO_lock_2023}, we have observed that TDDFT level of theory provides a great cost/accuracy tradeoff while still reproducing the essential features obtained by using higher-level, albeit much costlier, multireference methods.

Another possible source of error is the size of the QM subsystem, the choice of which involves a delicate balance of computational feasibility and accuracy. Previous work done by our group\cite{CO_lock_2023,gonza_jctc_2024_dftb_naf} has employed QM dimers for a variety of amino acid crystals and found that these reasonably faithfully reproduce the key experimentally observed behavior. In the case of cysteine, there still remains a noticeable blue-shift of the absorption spectra and the lowest excitation energy. In order to understand how sensitive the absorption spectra are to the size of the QM region, we performed some additional benchmarks. Indeed, SI Figure \ref{fig:SI_abs_validation}d demonstrates that increasing the QM region from a dimer to a heptamer (which contains 98 atoms) results in a red-shift of the spectra and subsequently reduces the discrepancy between the theory and experiment to $\sim$ \SI{2.1}{\electronvolt} while preserving the same physical origin for the transition to the first excited state. 

Finally, we wanted to verify the impact of both a larger QM region (SI Figure \ref{fig:SI_CI_validation}a), as well as a different electronic structure method (ADC(2)) near the conical intersection (SI Figure \ref{fig:SI_CI_validation}b). ADC(2) is widely considered the most accurate single-reference method for describing conical intersections \cite{plasser2014surface,papineau2024electronic}. As previously mentioned, in our TDDFT/TSH setup, we set the criterion for a hop to the ground state when the energy gap falls below \SI{0.2}{\electronvolt}. In the validation using a larger QM region, we see that the energy gaps between the first excited state and the ground state are preserved remarkably well when using pentamers instead of dimers as the QM region. In the second test, we applied ADC(2) using the same QM dimer as in the TDDFT case.  While some differences emerge compared to TDDFT, the overall energy gaps and conical intersection remain largely unchanged.


\begin{acknowledgement}
DB, GDM, MM, and AH thank the European Commission for funding on the ERC Grant HyBOP 101043272. DB, GDM, MM, and AH also acknowledge CINECA supercomputing (project NAFAA-HP10B4ZBB2) and MareNostrum5 (project EHPC-EXT-2023E01-029) for computational resources.
\end{acknowledgement}

\clearpage

\begin{suppinfo}
\beginsupplement
\subsection*{Data collection for the Cysteine crystal}
A Cys crystal suitable for X-ray diffraction was immersed in Paratone--N oil and mounted in a MiTeGen loop and flash-frozen in liquid nitrogen at \SI{120}{\kelvin}. Data were collected with a Rigaku XtaLabPro with microfocus sealed tube source and a Dectris Pilatus 3R 200K-A detector using CuK$\alpha$ radiation (\SI{1.54184}{\angstrom}). Data were collected as $\omega$ scans using CrysAlisPro 1.171.42.67a suite of programs (Rigaku OD, 2022). Data were integrated and reduced in CrysAlisPro. The crystal structure was solved by using direct methods (SHELXT-2016/4)\citep{sheldrick2015shelxt} and refined by full-matrix least-squares methods against $F^2$ (SHELXL-2016/4)\citep{sheldrick2015crystal} as implemented in Olex2\citep{dolomanov2009olex2}. All the non-hydrogen atoms were refined with anisotropic displacement parameters. The hydrogen atoms were located in the electron density and refined isotropically. The Mercury 2020.3.0 software was used for molecular graphics.

\newpage

\begin{table}[H]
\caption{Crystallographic data collection and refinement statistics data for the crystal of Cysteine in \ce{D2O} at \SI{120}{\kelvin}.}
\label{tab:SI_crystal_data}
\begin{tabular}{ll}
\hline
Compound & Cysteine (\ce{D2O}) \\
CCDC number & 2358399 \\
Crystal description & Colorless Needle \\
Diffractometer & Rigaku XtaLabPro \\
Temperature (K) & 120 \\
Chemical Formula & \ce{C6H14N2O4S2} \\
Formula weight (g/mol) & 242.31 \\
Radiation & CuK$\alpha$ ($\lambda = \SI{1.54184}{\angstrom}$) \\
$z$ & 2 \\
Crystal System & Orthorhombic \\
Bond-length (C-C) (\si{\angstrom}) & 0.0035 \\
Space group & P2$_1$2$_1$2$_1$ \\
$a$ (\si{\angstrom}) & 5.41839(11) \\
$b$ (\si{\angstrom}) & 8.13325(17) \\
$c$ (\si{\angstrom}) & 12.0157(2) \\
$\alpha$ (\si{\degree}) & 90 \\
$\beta$ (\si{\degree}) & 90 \\
$\gamma$ (\si{\degree}) & 90 \\
Volume (\si{\angstrom^3}) & 529.520(18) \\
$\rho$ (g/cm$^3$) & 1.520 \\
$\mu$ (mm$^{-1}$) & 4.552 \\
$F(000)$ & 256.0 \\
Crystal size/mm$^3$ & 0.355 × 0.077 × 0.073 \\
2$\Theta$ range for data collection/\si{\degree} & 13.144 to 158.22 \\
Index ranges & $-3 \le h \le 6$, $-10 \le k \le 10$, $-14 \le l \le 15$ \\
Reflections collected & 2622 \\
Independent reflections & 1128 [$R_\text{int}$ = 0.0283, $R_\text{sigma}$ = 0.0311] \\
Data/restraints/parameters & 1128/0/92 \\
Goodness-of-fit on $F^2$ & 1.120 \\
Final R indexes [I$ \ge 2\sigma$ (I)] & $R_1$ = 0.0336, w$R_2$ = 0.0914 \\
Final R indexes [all data] & $R_1$ = 0.0337, w$R_2$ = 0.0914 \\
Largest diff. peak/hole / e \si{\angstrom^{-3}} & 0.28/-0.43 \\
Flack parameter & 0.02(2) \\
\hline
\end{tabular}
\end{table}

\newpage

\subsection*{Details of excitation and non-radiative decay modes}
\begin{center}
\begin{table}[H]
\centering
\caption{A summary of the nature of $\text{S}_0 \rightarrow \text{S}_1$ excitations for Cys-H and Cys-D.}
\begin{tabular}{ccc}
\toprule
Nature of $S_0 \rightarrow S_1$ excitation & Cys-H & Cys-D \\
\midrule
$n \rightarrow \pi*$ (\ce{C-O}) & 23 & 29 \\
$n \rightarrow \sigma*$ (\ce{S-H}) & 17 & 11 \\
\bottomrule
\end{tabular}
\label{tab:excitations}
\end{table}


\begin{table}[H]
\centering
\caption{A summary of the different modes of non-radiative decay for Cys-H and Cys-D.}
\label{tab:decay_modes}
\begin{tabular}{ccc}
\toprule
Non-radiative decay mode & Cys-H & Cys-D \\
\midrule
\ce{S-H}                     & 12           & 9            \\
\ce{C_\alpha-N}                      & 19           & -            \\
\ce{C_\beta-S}                      & 4            & 1            \\
\ce{C_\alpha-COO-}              & -            & 3            \\
\bottomrule
\end{tabular}
\end{table}

\vspace*{\fill}
\end{center}

\newpage
\begin{center}
\vspace*{\fill}
\begin{figure}[ht!]
\centering
\includegraphics[width=\linewidth]{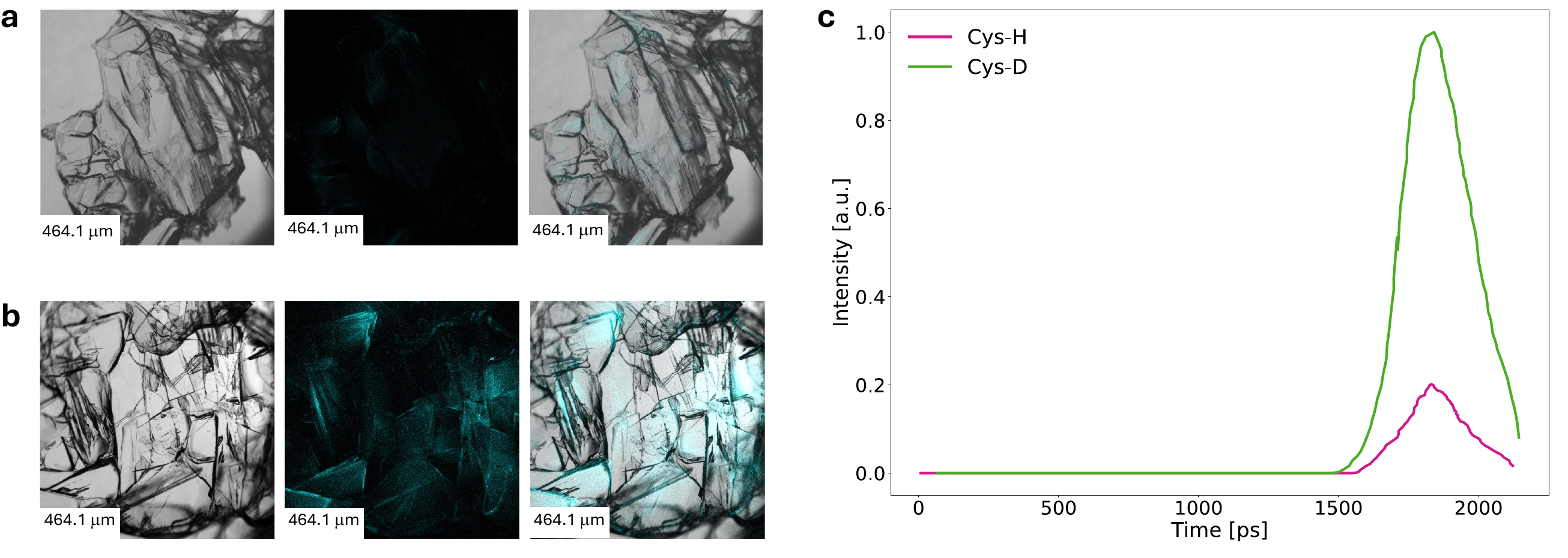}
\caption{In panels \textbf{a} and \textbf{b}, the confocal microscopic images of aggregate of crystals of Cys-H and Cys-D that are excited at a wavelength of \SI{405}{\nano\meter} are shown, respectively. Each of these two panels consist of brightfield, fluorescence, and merged image of the crystals (from left to right). In panel \textbf{c}, the FLIM spectra (excited at \SI{480}{\nano\meter}) of crystal aggregates captured with a time-correlated single-photon counting (TCSPC) system for Cys-H and Cys-D is shown in pink and green, respectively.}
\label{fig:SI_expt_flim_agg}
\end{figure}
\vspace*{\fill}
\end{center}

\newpage
\begin{center}
\vspace*{\fill}
\begin{figure}[ht!]
\centering
\includegraphics[width=\linewidth]{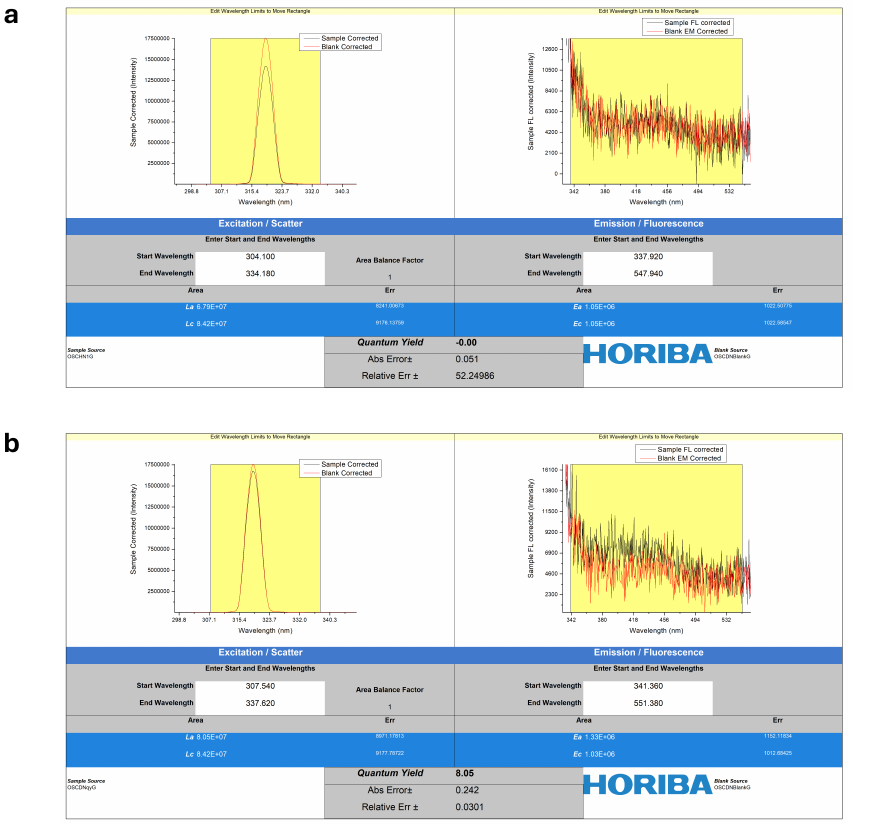}
\caption{Quantum yield of Cys-H (\textbf{a}) and Cys-D (\textbf{b}) measured under excitation of 320 nm for multiple crystals.}
\label{fig:SI_quantum_yield}
\end{figure}
\vspace*{\fill}
\end{center}


\newpage
\begin{center}
\vspace*{\fill}
\begin{figure}[ht!]
\centering
\includegraphics[width=0.8\linewidth]{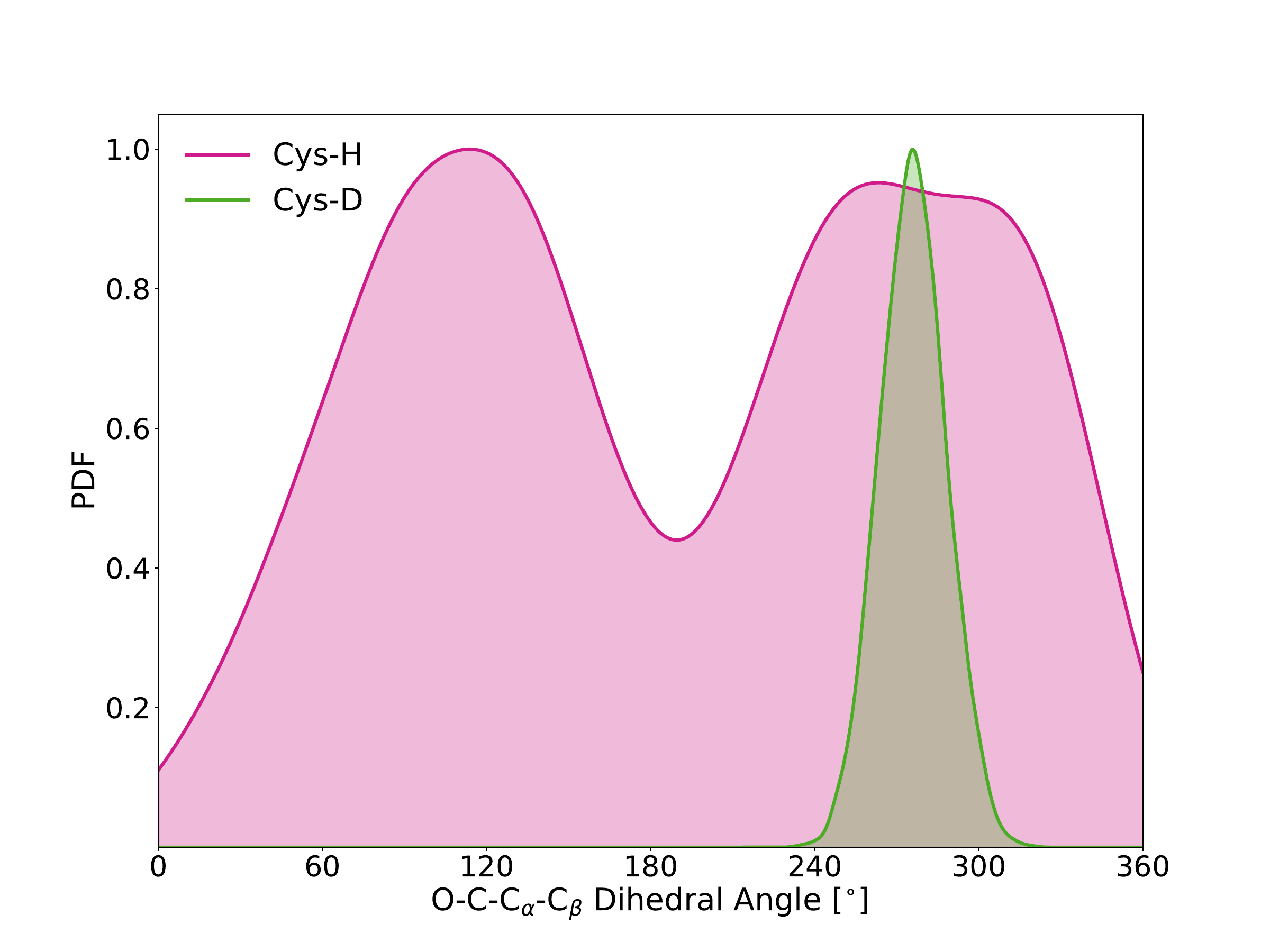}
\caption{A distribution of the \ce{O-C-C_\alpha-C_\beta} dihedral angle for both Cys-H (pink) and Cys-D (green) from finite-temperature classical NVT simulations in the ground state.}
\label{fig:SI_OCCaCb_GS}
\end{figure}
\vspace*{\fill}
\end{center}

\newpage
\begin{center}
\vspace*{\fill}
\begin{figure}[ht!]
\centering
\includegraphics[width=0.9\linewidth]{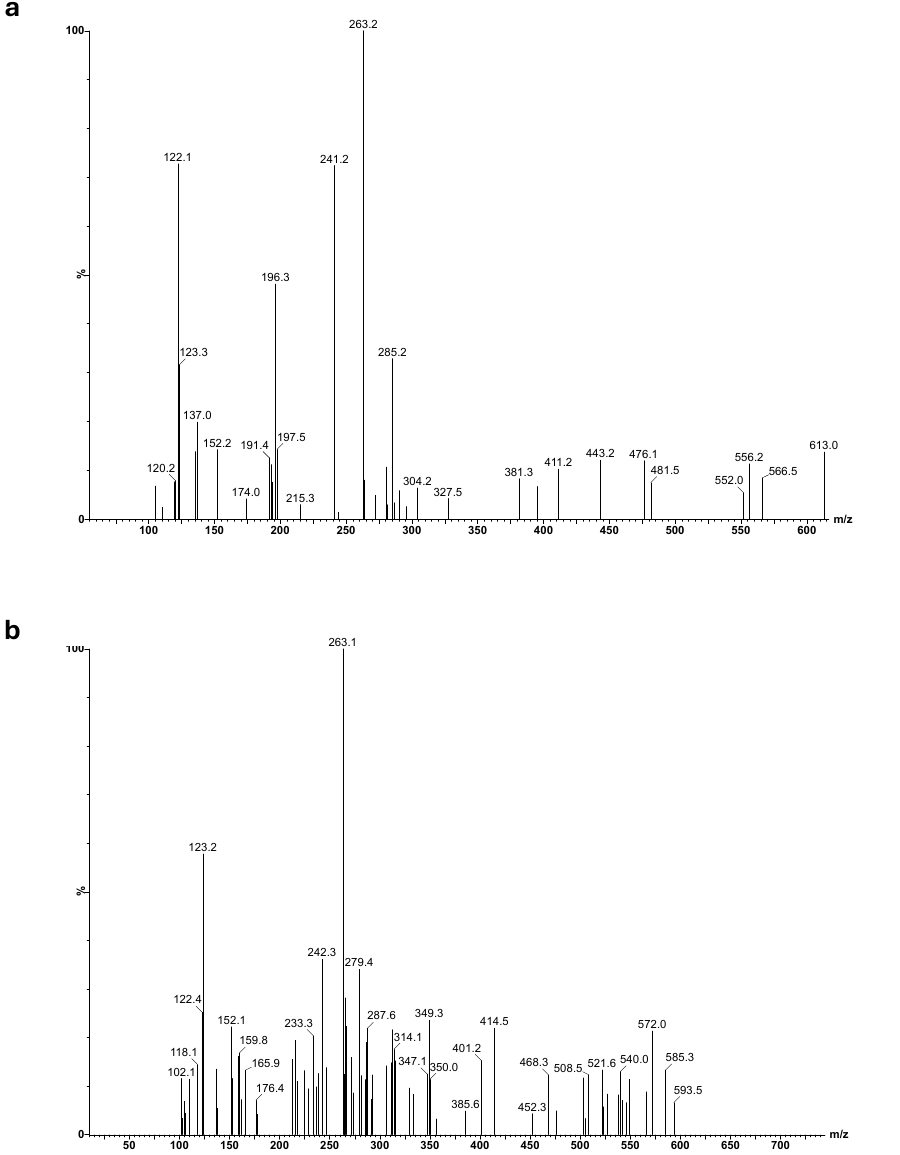}
\caption{Mass spectroscopy for Cys-H (\textbf{a}) and Cys-D (\textbf{b}) in positive-ion mode.}
\label{fig:SI_mass_spec}
\end{figure}
\vspace*{\fill}
\end{center}

\newpage
\begin{center}
\vspace*{\fill}
\begin{figure}[ht!]
    \centering
    \includegraphics[width=\linewidth]{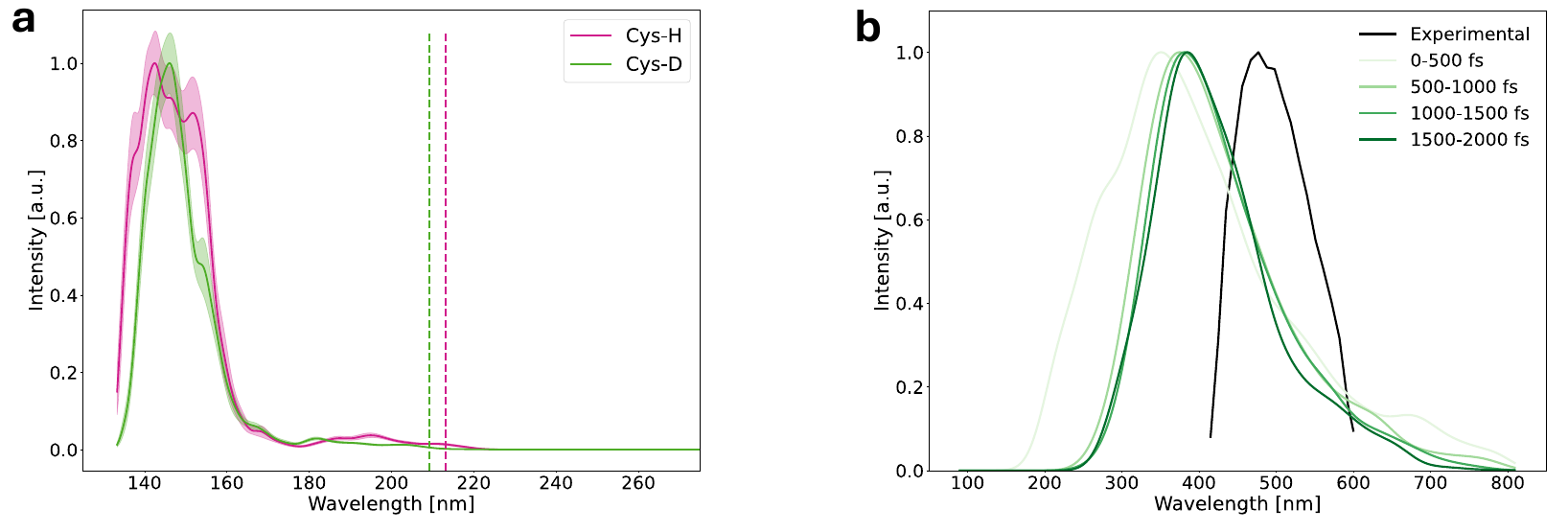}
    \caption{Panel \textbf{a} shows the theoretically computed absorption spectra for Cys-H and Cys-D in pink and green, respectively. The shaded region denotes error bars over 40 independent estimates. The dashed vertical lines denote positions of the lowest energy excitations ($\text{S}_0 \rightarrow \text{S}_1$) which are \SI{215}{\nano\meter} and \SI{209}{\nano\meter} for Cys-H and Cys-D, respectively. Panel \textbf{b} shows the time-dependent emission spectra from $\text{S}_1$ for Cys-D, in blocks of 500 fs, with the shades of green getting darker for successive blocks of time, while the experimental emission spectra is shown in black.}
    \label{fig:SI_spectra_theo}
\end{figure}
\vspace*{\fill}
\end{center}

\newpage
\begin{center}
\vspace*{\fill}
\begin{figure}[ht!]
    \centering
    \includegraphics[width=\linewidth]{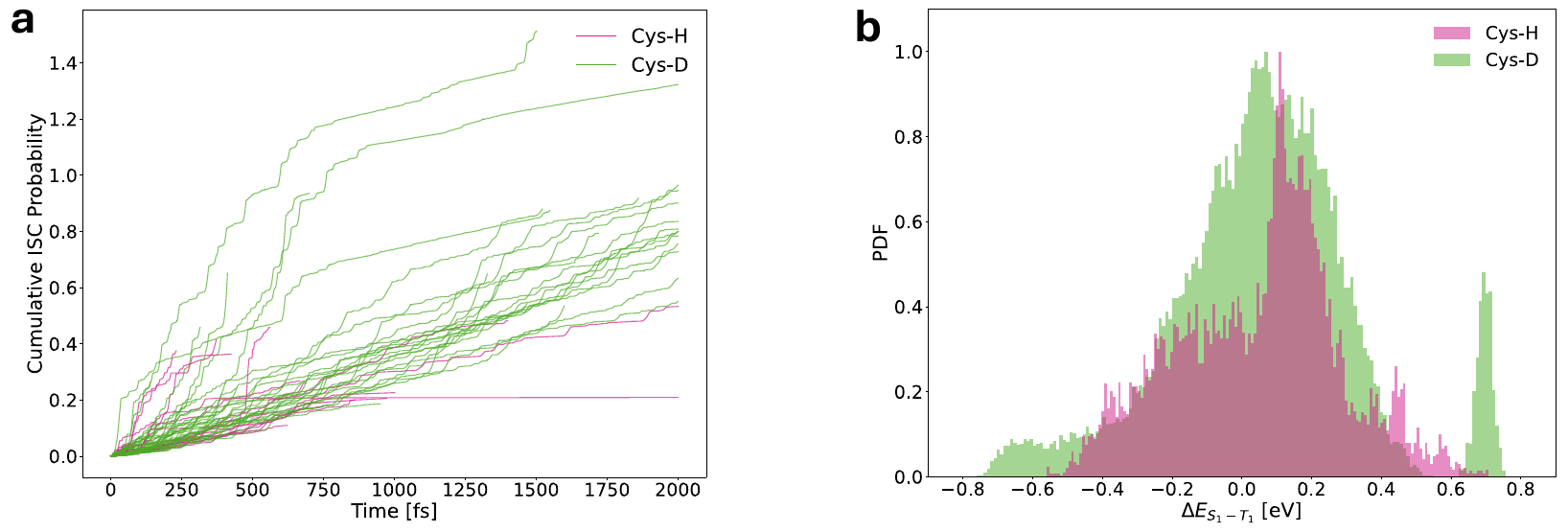}
    \caption{Panel \textbf{a} shows the time-evolution of the cumulative ISC probability from $\text{S}_1$ to any of the triplet states for each NAMD trajectory. Panel \textbf{b} shows a distribution of the energy gaps between $\text{S}_1$ and $\text{T}_x$ (which denotes the energetically closest triplet state, usually $\text{T}_1$). The data for Cys-H and Cys-D is shown in pink and green, respectively.}
    \label{fig:SI_isc_prob_energy-gaps}
\end{figure}
\vspace*{\fill}
\end{center}


\newpage
\begin{center}
\vspace*{\fill}
\begin{figure}[ht!]
    \centering
    \includegraphics[width=\linewidth]{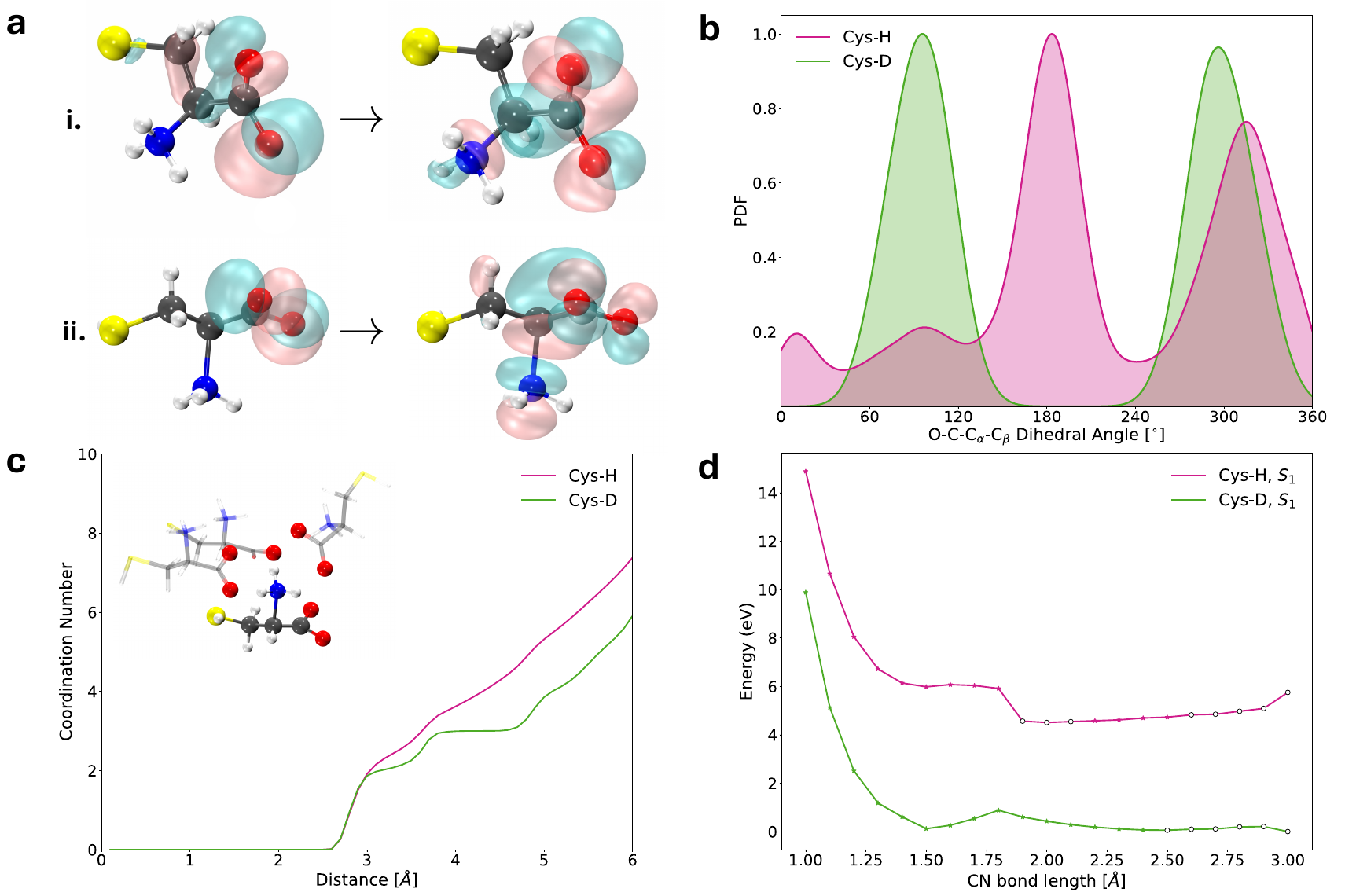}
    \caption{
    Panel \textbf{a.i} shows the occupied (left) and virtual (right) orbitals for the $\text{S}_0 \rightarrow \text{S}_1$ transition at the starting frame of a reference NAMD trajectory where Cys-H decays non-radiatively to $\text{S}_0$ using the C-N stretching mode. Panel \textbf{a.ii} shows the same orbitals, for the same NAMD trajectory, at the frame of encountering the conical intersection. \\
    Panel \textbf{b} shows a distribution of the \ce{O-C-C_\alpha-C_\beta} dihedral angle for both Cys-H (pink) and Cys-D (green) while the system is in $\text{S}_1$. \\
    Panel \textbf{c} displays the running coordination number of Nitrogen-Oxygen for Cys-H (pink) and Cys-D (green). The inset of the figure shows a sample frame from an NAMD trajectory where the it is clear that the \ce{NH3+} stretches into a ``cavity" of O atoms.\\
    Panel \textbf{d} shows the excited state relaxed potential energy scan for different values of the \ce{N-C_\alpha} bond with the lines showing the energies for the optimized structures in $\text{S}_1$ for each bond length scan value, for Cys-H (pink) and Cys-D (green). The small circles represent points where the geometry optimization has not converged even after 1500 optimization cycles, which are usually those very close to the conical intersection. For clarity of visualization, the pink curve corresponding to Cys-H is shifted up by \SI{5}{\electronvolt}.}
    \label{fig:SI_CN_analysis}
\end{figure}
\vspace*{\fill}
\end{center}

\newpage
\begin{center}
\vspace*{\fill}
\begin{figure}[ht!]
    \centering
    \includegraphics[width=\linewidth]{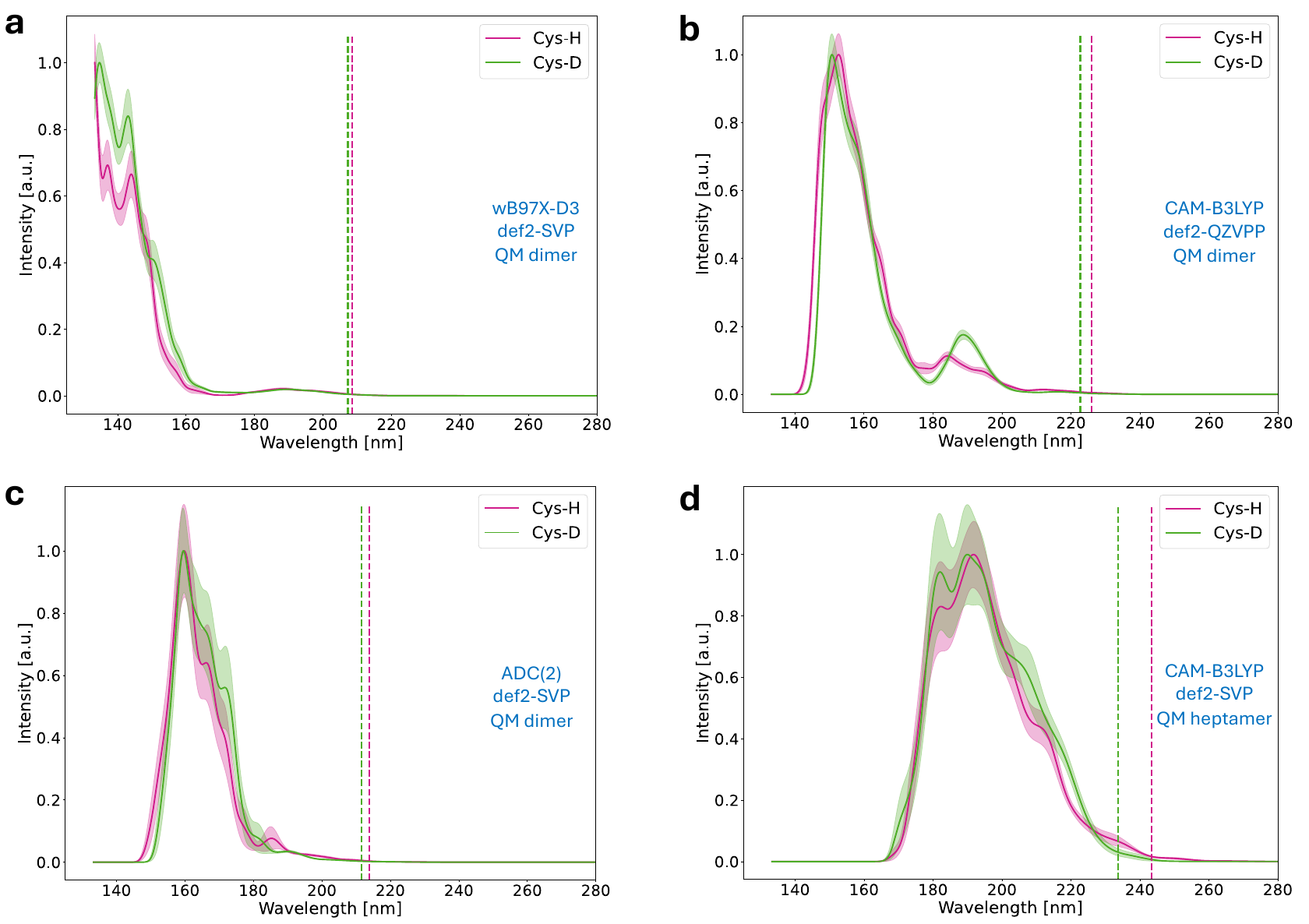}
    \caption{
    This figure shows the results of a variety of theoretical absorption spectra calculations to test the accuracy of our choice of electronic structure and QM/MM setup. In all panels the shaded curves in pink and green correspond to Cys-H and Cys-D respectively, while the dashed vertical lines correspond to the lowest energy excitations ($\text{S}_0 \rightarrow \text{S}_1$), following the same colorscheme.\\
    Panel \textbf{a}: TD-DFT level of theory using a dimer QM region with wB97X-D3 functional and def2-SVP basis set. \\
    Panel \textbf{b}: TD-DFT level of theory using a dimer QM region with CAM-B3LYP functional and def2-QZVPP basis set. \\
    Panel \textbf{c}: ADC(2) level of theory using a dimer QM region with def2-SVP basis set. \\
    Panel \textbf{d}: TD-DFT level of theory using a heptamer QM region (98 atoms) with CAM-B3LYP functional and def2-SVP basis set. \\
    In all cases the calculations are done using ORCA 5.0.4.
    }
    \label{fig:SI_abs_validation}
\end{figure}
\vspace*{\fill}
\end{center}

\newpage
\begin{center}
\vspace*{\fill}
\begin{figure}[ht!]
    \centering
    \includegraphics[width=\linewidth]{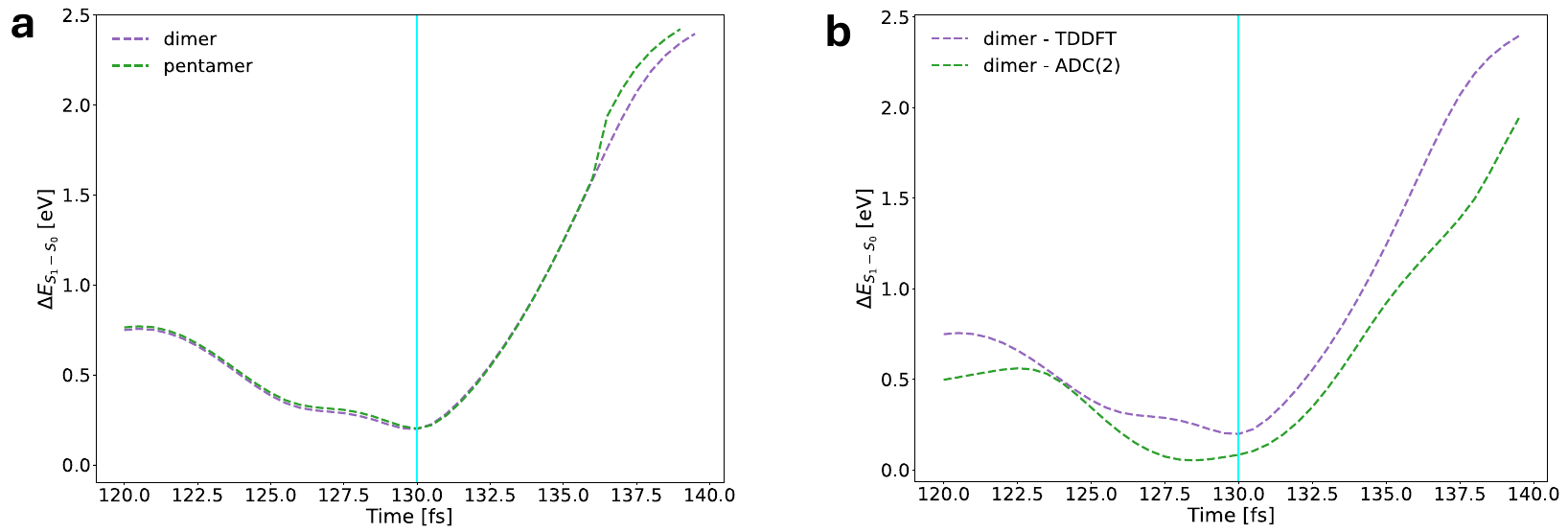}
    \caption{
    Panel \textbf{a} shows the energy gap between $\text{S}_1$ and $\text{S}_0$ for a Cys-H trajectory decaying via the \ce{C-N} mode, near the conical intersection, where the original energy gap considering dimers in the QM region is shown with purple dashed lines, and the recomputed energy gap expanding the QM region to a pentamer is shown in green dashed lines. These are both computed using TDDFT.\\
    Panel \textbf{b} shows the same energy gap for the QM dimer as in \textbf{a}, computed at both TDDFT (purple) and ADC(2) (green) levels of theory using ORCA 5.0.4.\\
    Here it should be noted that while attempting the validation of the pentamer energies using ADC(2), as implemented in ORCA 5.0.4, we ran into issues pertaining to computational cost.\\
    We reiterate here that since TDDFT has issues describing the CI between $\text{S}_1$ and $\text{S}_0$, we choose a cutoff value of \SI{0.2}{\electronvolt} to determine a hop to the ground state.}
    \label{fig:SI_CI_validation}
\end{figure}
\vspace*{\fill}
\end{center}

\end{suppinfo}

\bibliography{refs}

\begin{tocentry}

\includegraphics[width=\linewidth]{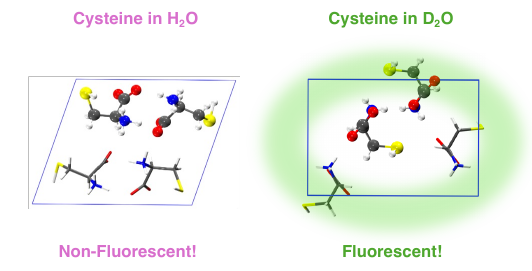}

\end{tocentry}

\end{document}